\documentclass[twocolumn,showpacs,preprintnumbers]{revtex4}%
\usepackage{amssymb}
\usepackage{amsmath}
\usepackage{graphicx}
\usepackage{dcolumn}
\usepackage{bm}
\usepackage{amsfonts}%
\begin{document}
\title{Decoherence Suppression by Cavity Optomechanical Cooling}
\author{Eyal Buks}
\affiliation{Department of Electrical Engineering, Technion, Haifa 32000 Israel}
\date{\today }

\begin{abstract}
We consider a cavity optomechanical cooling configuration consisting of a
mechanical resonator (denoted as resonator $b$) and an electromagnetic
resonator (denoted as resonator $a$), which are coupled in such a way that the
effective resonance frequency of resonator $a$ depends linearly on the
displacement of resonator $b$. We study whether back-reaction effects in such
a configuration can be efficiently employed for suppression of decoherence. To
that end, we consider the case where the mechanical resonator is prepared in a
superposition of two coherent states and evaluate the rate of decoherence. We
find that no significant suppression of decoherence is achievable when
resonator $a$ is assumed to have a linear response. On the other hand, when
resonator $a$ exhibits Kerr nonlinearity and/or nonlinear damping the
decoherence rate can be made much smaller than the equilibrium value provided
that the parameters that characterize these nonlinearities can be tuned close
to some specified optimum values.

\end{abstract}
\maketitle





\section{Introduction}

The quest for quantum effects in nanomechanical devices has motivated an
intense research effort in recent years
\cite{Blencowe_159,Schwab_36,OConnell_697}. Experimental demonstration of
quantum superposition in a nanomechanical resonator may provide an important
insight into the problem of quantum to classical transition
\cite{Penrose_581,Diosi_1165,Legget_R415,Leggett_857,Bose_4175,Bose_3204,Kleckner_095020}%
. However, in many cases the lifetime of such superposition states is too
short for experimental observation since the coupling between a nanomechanical
resonator and its environment typically results in rapid decoherence
\cite{Zurek_0306072,Zurek_715}. As a case study, consider a superposition of
two coherent states $\left\vert \alpha_{1}\right\rangle $ and $\left\vert
\alpha_{2}\right\rangle $ of a mechanical resonator having an angular
resonance frequency $\omega_{b}$ and damping rate $\gamma_{b}$. The
decoherence rate of such a superposition state is given in the high
temperature limit $k_{\mathrm{B}}T\gg\hbar\omega_{b}$ by \cite{Caldeira_587,
Joos_223, Unruh_1071, Zurek_36}%
\begin{equation}
\frac{1}{\tau_{\varphi}}=4\gamma_{b}\left\vert \delta_{\alpha}\right\vert
^{2}\frac{k_{\mathrm{B}}T}{\hbar\omega_{b}}\;, \label{1/tau TE}%
\end{equation}
where $\delta_{\alpha}=\alpha_{2}-\alpha_{1}$.

While Eq. (\ref{1/tau TE}) was derived by assuming linear response, it is well
known that nonlinear response can be exploited for reduction of thermal
fluctuations. One example is the technique of noise squeezing that can be
employed for reducing thermal fluctuations in one of the quadratures of a
mechanical resonator \cite{Rugar_699,Almog_078103}. Another example, which is
the focus of this chapter, is the technique of optomechanical cavity cooling.
This technique
\cite{Braginsky_2002,Martin_125339,Wilson-Rae_075507,Clerk_238,Blencowe_236,Wineland_0606180,Marquardt_093902,Kimble_et_al_01}%
, which was first proposed as a way to enhance the detection sensitivity of
gravity waves \cite{Braginsky&Manukin_67, Braginsky_et_al_70}, can be employed
for significantly reducing the energy fluctuations of a mechanical resonator
well below the equilibrium value
\cite{Hohberger-Metzger_1002,Gigan_67,Arcizet_71,Kleckner_75,Corbitt_et_al_06,Corbitt_150802,Schliesser_243905,Harris_013107,Naik_193,Schliesser_et_al_08,
Genes_et_al_08, Kippenberg&Vahala_08, Teufel_et_al_10,Teufel_1103_2144}.
Cooling is achieved by coupling the mechanical resonator (denoted as resonator
$b$) to an electromagnetic resonator (denoted as resonator $a$) in such a way
that the effective resonance frequency of resonator $a$ becomes linearly
dependent on the displacement of resonator $b$. When the parameters of the
system are optimally chosen the fluctuations of resonator $b$ around steady
state can be significantly reduced well below the equilibrium value by
externally driving resonator $a$ with a monochromatic pump tone. In this
region back-reaction due to the retarded response of the driven resonator $a$
to fluctuations of resonator $b$ acts as a negative feedback, providing thus
additional damping which results in effective cooling down of resonator $b$.
The success of these experiments raises the question whether similar
back-reaction effects can also be efficiently employed for suppression of
decoherence below the equilibrium value.

Here we study this problem by generalizing Eq. (\ref{1/tau TE}) for the case
where cavity cooling is applied. Nonlinearity in resonator $a$ is taken into
account to lowest nonvanishing order. The equations of motion of the system
are obtained using the Gardiner and Collett input-output theory
\cite{Gardiner_3761,Yurke_5054}. By linearizing these equations we derive the
susceptibility matrixes of the system, which allow calculating the response of
both resonators to input noise. This, in turn, allows evaluating both, the
spectral density of fluctuations and the decoherence rate $1/\tau_{\varphi}$
of resonator $b$. In both cases we examine the cooling efficiency by defining
an appropriate effective temperature and by calculating it for an optimum
choice of the system's parameters. We find that only modest suppression of
decoherence is possible using cavity cooling unless the system is driven into
the region of nonlinear oscillations.

\section{The Model}

The model consists of two resonators, labeled as $a$ and $b$ respectively,
which are coupled to each other by a term $\hbar\Omega N_{a}\left(
A_{b}+A_{b}^{\dag}\right)  $ in the Hamiltonian. Here $A_{a}$, $A_{a}^{\dag}$
\ and $N_{a}=A_{a}^{\dag}A_{a}$ ($A_{b}$, $A_{b}^{\dag}$ and $N_{b}%
=A_{b}^{\dag}A_{b}$) are respectively annihilation, creation and number
operators of resonator $a$ ($b$). The first resonator is coupled to 3
semi-infinite transmission lines. The first, denoted as $a1$, is a feedline,
which is linearly coupled to resonator $a$ with a coupling constant $T_{a1}$,
and which is employed to deliver the input and output signals; the second,
denoted as $a2$, is linearly coupled to resonator $a$ with a coupling constant
$T_{a2}$, and it is used to model linear dissipation, whereas the third one,
denoted as $a3$, is nonlinearly coupled to resonator $a$ with a coupling
constant $T_{a3}$, and is employed to model nonlinear dissipation. Linear
dissipation of resonator $b$ is modeled using semi-infinite transmission line,
which is denoted as $b$ and which is linearly coupled to resonator $b$ with a
coupling constant $T_{b}$. Kerr-like nonlinearity of the driven resonator $a$
is taken into account to lowest order by including the term $\left(
\hbar/2\right)  K_{a}A_{a}^{\dagger}A_{a}^{\dagger}A_{a}A_{a}$ in the
Hamiltonian of the system, which is given by%
\begin{align}
\mathcal{H} &  =\hbar\omega_{a}N_{a}+\frac{\hbar}{2}K_{a}A_{a}^{\dagger}%
A_{a}^{\dagger}A_{a}A_{a}+\hbar\omega_{b}N_{b}\nonumber\\
&  +\hbar\Omega N_{a}\left(  A_{b}+A_{b}^{\dag}\right)  \nonumber\\
&  +\hbar\int\mathrm{d}\omega\;a_{a1}^{\dagger}\left(  \omega\right)
a_{a1}\left(  \omega\right)  \omega\nonumber\\
&  +\hbar\int\mathrm{d}\omega\;\left[  T_{a1}A_{a}^{\dagger}a_{a1}\left(
\omega\right)  +T_{a1}^{\ast}a_{a1}^{\dagger}\left(  \omega\right)
A_{a}\right]  \nonumber\\
&  +\hbar\int\mathrm{d}\omega\;a_{a2}^{\dagger}\left(  \omega\right)
a_{a2}\left(  \omega\right)  \omega\nonumber\\
&  +\hbar\int\mathrm{d}\omega\;\left[  T_{a2}A_{a}^{\dagger}a_{a2}\left(
\omega\right)  +T_{a2}^{\ast}a_{a2}^{\dagger}\left(  \omega\right)
A_{a}\right]  \nonumber\\
&  +\hbar\int\mathrm{d}\omega\;a_{a3}^{\dagger}\left(  \omega\right)
a_{a3}\left(  \omega\right)  \omega\nonumber\\
&  +\hbar\int\mathrm{d}\omega\;\left[  T_{a3}A_{a}^{\dagger}A_{a}^{\dagger
}a_{a3}\left(  \omega\right)  +T_{a3}^{\ast}a_{a3}^{\dagger}\left(
\omega\right)  A_{a}A_{a}\right]  \nonumber\\
&  +\hbar\int\mathrm{d}\omega\;a_{b}^{\dagger}\left(  \omega\right)
a_{b}\left(  \omega\right)  \omega\nonumber\\
&  +\hbar\int\mathrm{d}\omega\;\left[  T_{b}A_{b}^{\dagger}a_{b}\left(
\omega\right)  +T_{b}^{\ast}a_{b}^{\dagger}\left(  \omega\right)
A_{b}\right]  \;.\nonumber\\
& \label{Hamiltonian}%
\end{align}

\subsection{Equations of Motion}

The Heisenberg equations of motion are generated according to
\begin{equation}
i\hbar\frac{\mathrm{d}O}{\mathrm{d}t}=\left[  O,\mathcal{H}\right]  \;,
\end{equation}
where $O$ is an operator. Using the commutation relations
\begin{align}
\left[  A_{a},A_{a}^{\dagger}\right]   &  =\left[  A_{b},A_{b}^{\dagger
}\right]  =1\;,\\
\left[  A_{a},N_{a}\right]   &  =A_{a}\;,\\
\left[  A_{b},N_{b}\right]   &  =A_{b}\;,\\
\left[  A_{a},A_{a}^{\dagger}A_{a}^{\dagger}A_{a}A_{a}\right]   &
=2N_{a}A_{a}\;,
\end{align}
one has%
\begin{align}
\frac{\mathrm{d}A_{a}}{\mathrm{d}t} &  =-i\omega_{a}A_{a}-iK_{a}N_{a}%
A_{a}-i\Omega A_{a}\left(  A_{b}+A_{b}^{\dag}\right)  \nonumber\\
&  -i\int\mathrm{d}\omega\;T_{a1}a_{a1}\left(  \omega\right)  -i\int
\mathrm{d}\omega\;T_{a2}a_{a2}\left(  \omega\right)  \nonumber\\
&  -2i\int\mathrm{d}\omega\;T_{a3}A_{a}^{\dagger}a_{a3}\left(  \omega\right)
\;,\nonumber\\
& \label{dA_a/dt}%
\end{align}
and%
\begin{equation}
\frac{\mathrm{d}A_{b}}{\mathrm{d}t}=-i\omega_{b}A_{b}-i\Omega N_{a}%
-i\int\mathrm{d}\omega\;T_{b}a_{b}\left(  \omega\right)  \;.\label{dA_b/dt}%
\end{equation}
Using the bath modes commutation relations%
\begin{align}
\left[  a_{a1}\left(  \omega\right)  ,a_{a1}^{\dagger}\left(  \omega^{\prime
}\right)  \right]   &  =\delta\left(  \omega-\omega^{\prime}\right)  \;,\\
\left[  a_{a1}\left(  \omega\right)  ,a_{a1}\left(  \omega^{\prime}\right)
\right]   &  =0\;,
\end{align}
one obtains%
\begin{equation}
\frac{\mathrm{d}a_{a1}\left(  \omega\right)  }{\mathrm{d}t}=-i\omega
a_{a1}\left(  \omega\right)  -iT_{a1}^{\ast}A_{a}\;.
\end{equation}
Using initial condition $a_{a1}\left(  \omega,t_{0}\right)  $ one finds by
integration that%
\begin{align}
a_{a1}\left(  \omega,t\right)   &  =a_{a1}\left(  \omega,t_{0}\right)
e^{i\omega\left(  t_{0}-t\right)  }\nonumber\\
&  -iT_{a1}^{\ast}\int_{t_{0}}^{t}\mathrm{d}t^{\prime}\;A_{a}\left(
t^{\prime}\right)  e^{i\omega\left(  t^{\prime}-t\right)  }\;.\nonumber\\
& \label{a_a1}%
\end{align}
Next we integrate Eq. (\ref{a_a1}) over $\omega$. The coupling coefficient
$T_{a1}$, which is assumed to be $\omega$ independent, is expressed as%
\begin{equation}
T_{a1}=\sqrt{\frac{\gamma_{a1}}{\pi}}e^{i\phi_{a1}}\;,
\end{equation}
where $\gamma_{a1}$ is positive and $\phi_{a1}$ is real. Using the following
relations%
\begin{equation}
\int\mathrm{d}\omega\;e^{i\omega\left(  t^{\prime}-t\right)  }=2\pi
\delta\left(  t-t^{\prime}\right)  \;,
\end{equation}%
\begin{equation}
\int_{t_{0}}^{t}\mathrm{d}t^{\prime}\;\delta\left(  t-t^{\prime}\right)
f\left(  t^{\prime}\right)  =\frac{1}{2}\mathrm{sgn}\left(  t-t_{0}\right)
f\left(  t\right)  \ ,
\end{equation}
where $\mathrm{sgn}(x)$ is the sign function%
\begin{equation}
\mathrm{sgn}(x)=\left\{
\begin{array}
[c]{cc}%
+1 & \mathrm{if}\ x>0\\
-1 & \mathrm{if}\ x<0.
\end{array}
\right.  \ ,
\end{equation}
one finds that%
\begin{equation}
\frac{1}{\sqrt{2\pi}}\int\mathrm{d}\omega\;a_{a1}\left(  \omega,t\right)
=a_{a1}^{\mathrm{in}}\left(  t\right)  -i\sqrt{\frac{\gamma_{a1}}{2}}%
e^{-i\phi_{a1}}A_{a}\left(  t\right)  \;,\;
\end{equation}
where%
\begin{equation}
a_{a1}^{\mathrm{in}}\left(  t\right)  =\frac{1}{\sqrt{2\pi}}\int
\mathrm{d}\omega\;a_{a1}\left(  \omega,t_{0}\right)  e^{i\omega\left(
t_{0}-t\right)  }\;.\label{a^in}%
\end{equation}

Using similar definitions the above results are generalized for the other
semi-infinite transmission lines that are linearly coupled (labeled as $a2$
and $b$). For the transmission line $a3$, which is nonlinearly coupled, the
coupling coefficient $T_{a3}$, which is also assumed to be $\omega$
independent, is expressed as%
\begin{equation}
T_{a3}=\sqrt{\frac{\gamma_{a3}}{2\pi}}e^{i\phi_{a3}}\;,
\end{equation}
and the following holds%
\begin{equation}
\frac{1}{\sqrt{2\pi}}\int\mathrm{d}\omega\;a_{a3}\left(  \omega,t\right)
=a_{a3}^{\mathrm{in}}\left(  t\right)  -i\frac{\sqrt{\gamma_{a3}}}{2}%
e^{-i\phi_{a3}}A_{a}\left(  t\right)  A_{a}\left(  t\right)  \;.
\end{equation}

Substituting these results into Eqs. (\ref{dA_a/dt}) and (\ref{dA_b/dt})
yields%
\begin{align}
\frac{\mathrm{d}A_{a}}{\mathrm{d}t}  &  =-\left[  i\omega_{a}+\gamma
_{a}+\left(  iK_{a}+\gamma_{a3}\right)  N_{a}\right]  A_{a}\nonumber\\
&  -i\Omega A_{a}\left(  A_{b}+A_{b}^{\dag}\right) \nonumber\\
&  -i\sqrt{2\gamma_{a1}}e^{i\phi_{a1}}a_{a1}^{\mathrm{in}}\left(  t\right)
-i\sqrt{2\gamma_{a2}}e^{i\phi_{a2}}a_{a2}^{\mathrm{in}}\left(  t\right)
\nonumber\\
&  -2i\sqrt{\gamma_{a3}}e^{i\phi_{a3}}A_{a}^{\dagger}a_{a3}^{\mathrm{in}%
}\left(  t\right)  \;,\nonumber\\
&  \label{dA_a/dt 2}%
\end{align}
and%
\begin{align}
\frac{\mathrm{d}A_{b}}{\mathrm{d}t}  &  =-\left(  i\omega_{b}+\gamma
_{b}\right)  A_{b}-i\Omega N_{a}\nonumber\\
&  -i\sqrt{2\gamma_{b}}e^{i\phi_{b}}a_{b}^{\mathrm{in}}\left(  t\right)
\;,\nonumber\\
&  \label{dA_b/dt 2}%
\end{align}
where%
\begin{equation}
\gamma_{a}=\gamma_{a1}+\gamma_{a2}\;.
\end{equation}

\subsection{Rotating Frame}

Consider the case where a coherent tone at angular frequency $\omega
_{\mathrm{p}}$ and a constant complex amplitude $b_{\mathrm{p}}$ is injected
into the feedline. The operators of the driven resonator and its thermal baths
are expressed in a frame rotating at frequency $\omega_{\mathrm{p}}$ as%
\begin{align}
a_{a1}^{\mathrm{in}}  &  =b_{\mathrm{p}}e^{-i\omega_{\mathrm{p}}t}%
+c_{a1}^{\mathrm{in}}e^{-i\omega_{\mathrm{p}}t}\;,\label{a_a1^in rot}\\
a_{a2}^{\mathrm{in}}  &  =c_{a2}^{\mathrm{in}}e^{-i\omega_{\mathrm{p}}%
t}\;,\label{a_a2^in rot}\\
a_{a3}^{\mathrm{in}}  &  =c_{a3}^{\mathrm{in}}e^{-i\omega_{\mathrm{p}}%
t}\;,\label{a_a3^in rot}\\
A_{a}  &  =C_{a}e^{-i\omega_{\mathrm{p}}t}\;, \label{A_a rot}%
\end{align}
Using this notation Eqs. (\ref{dA_a/dt 2}) and (\ref{dA_b/dt 2}) can be
rewritten as%
\begin{equation}
\frac{\mathrm{d}C_{a}}{\mathrm{d}t}+\Theta_{a}=F_{a}\;, \label{dC_a/dt r}%
\end{equation}%
\begin{equation}
\frac{\mathrm{d}A_{b}}{\mathrm{d}t}+\Theta_{b}=F_{b}\;, \label{dA_b/dt r}%
\end{equation}
where%
\begin{align}
\Theta_{a}  &  =\Theta_{a}\left(  C_{a},C_{a}^{\dag},A_{b},A_{b}^{\dag}\right)
\nonumber\\
&  =\left\{  i\left[  \Delta_{a}+\Omega\left(  A_{b}+A_{b}^{\dag}\right)
\right]  +\gamma_{a}+\left(  iK_{a}+\gamma_{a3}\right)  N_{a}\right\}
C_{a}\nonumber\\
&  +i\sqrt{2\gamma_{a1}}e^{i\phi_{a1}}b_{\mathrm{p}}\;,\nonumber\\
&
\end{align}%
\begin{equation}
\Delta_{a}=\omega_{a}-\omega_{\mathrm{p}}\;,
\end{equation}%
\begin{align}
F_{a}  &  =-i\sqrt{2\gamma_{a1}}e^{i\phi_{a1}}c_{a1}^{\mathrm{in}}%
-i\sqrt{2\gamma_{a2}}e^{i\phi_{a2}}c_{a2}^{\mathrm{in}}\nonumber\\
&  -2i\sqrt{\gamma_{a3}}e^{i\left(  \phi_{a3}+\omega_{\mathrm{p}}t\right)
}C_{a}^{\dag}c_{a3}^{\mathrm{in}}\;,\nonumber\\
&  \label{F_a}%
\end{align}%
\begin{align}
\Theta_{b}  &  =\Theta_{b}\left(  C_{a},C_{a}^{\dag},A_{b},A_{b}^{\dag}\right)
\nonumber\\
&  =\left(  i\omega_{b}+\gamma_{b}\right)  A_{b}+i\Omega N_{a}\;\nonumber\\
&
\end{align}
and%
\begin{equation}
F_{b}=-i\sqrt{2\gamma_{b}}e^{i\phi_{b}}a_{b}^{\mathrm{in}}\left(  t\right)
\;. \label{F_b}%
\end{equation}

\section{Linearization}

Expressing the solution as%
\begin{subequations}
\begin{align}
C_{a} &  =B_{a}+c_{a}\;,\\
A_{b} &  =B_{b}+c_{b}\;,
\end{align}
where both $B_{a}$ and $B_{b}$ are complex numbers, and considering both
$c_{a}$ and $c_{b}$ as small one has to lowest order%
\end{subequations}
\begin{subequations}
\begin{align}
\Theta_{a}\left(  C_{a},C_{a}^{\dag},C_{b},C_{b}^{\dag}\right)   &
=\Theta_{a}\left(  B_{a},B_{a}^{\ast},B_{b},B_{b}^{\ast}\right)  \nonumber\\
&  +W_{1}c_{a}+W_{2}c_{a}^{\dag}+W_{3}c_{b}+W_{4}c_{b}^{\dag}\;,\nonumber\\
& \label{Theta_a expansion}%
\end{align}%
\end{subequations}
\begin{subequations}
\begin{align}
\Theta_{b}\left(  C_{a},C_{a}^{\dag},C_{b},C_{b}^{\dag}\right)   &
=\Theta_{b}\left(  B_{a},B_{a}^{\ast},B_{b},B_{b}^{\ast}\right)  \nonumber\\
&  +W_{5}c_{a}+W_{6}c_{a}^{\dag}+W_{7}c_{b}+W_{8}c_{b}^{\dag}\;,\nonumber\\
& \label{Theta_b expansion}%
\end{align}
where%
\end{subequations}
\begin{subequations}
\begin{align}
W_{1} &  =i\Delta_{a}^{\mathrm{eff}}+\gamma_{a}+2\left(  iK_{a}+\gamma
_{a3}\right)  \left\vert B_{a}\right\vert ^{2}\;,\label{W1}\\
W_{2} &  =\left(  iK_{a}+\gamma_{a3}\right)  B_{a}^{2}\;,\\
W_{3} &  =W_{4}=i\Omega B_{a}\;,\\
W_{5} &  =i\Omega B_{a}^{\ast}\;,\\
W_{6} &  =i\Omega B_{a}\;,\\
W_{7} &  =i\omega_{b}+\gamma_{b}\;,\\
W_{8} &  =0\;,
\end{align}
and where%
\end{subequations}
\begin{equation}
\Delta_{a}^{\mathrm{eff}}=\Delta_{a}+\Omega\left(  B_{b}+B_{b}^{\ast}\right)
\;.
\end{equation}

\subsection{Mean Field Solution}

Mean field solutions are found by solving%
\begin{subequations}
\begin{align}
\Theta_{a}\left(  B_{a},B_{a}^{\ast},B_{b},B_{b}^{\ast}\right)   &  =0\;,\\
\Theta_{b}\left(  B_{a},B_{a}^{\ast},B_{b},B_{b}^{\ast}\right)   &  =0\;,
\end{align}
that is%
\end{subequations}
\begin{align}
&  \left[  i\Delta_{a}^{\mathrm{eff}}+\gamma_{a}+\left(  iK_{a}+\gamma
_{a3}\right)  \left\vert B_{a}\right\vert ^{2}\right]  B_{a}\nonumber\\
&  +i\sqrt{2\gamma_{a1}}e^{i\phi_{a1}}b_{\mathrm{p}}=0\;\nonumber\\
& \label{Theta_a=0}%
\end{align}
and%
\begin{equation}
\left(  i\omega_{b}+\gamma_{b}\right)  B_{b}+i\Omega\left\vert B_{a}%
\right\vert ^{2}=0\;.\label{Theta_b=0}%
\end{equation}
Extracting $B_{b}$ from Eq. (\ref{Theta_b=0}) and substituting it in Eq.
(\ref{Theta_a=0}) yields%
\begin{align}
&  \left\{  i\Delta_{a}+\gamma_{a}+\left(  iK_{a}^{\mathrm{eff}}+\gamma
_{a3}\right)  \left\vert B_{a}\right\vert ^{2}\right\}  B_{a}\nonumber\\
&  +i\sqrt{2\gamma_{a1}}e^{i\phi_{a1}}b_{\mathrm{p}}=0\;,\nonumber\\
& \label{Eq. for B_a}%
\end{align}
where $K_{a}^{\mathrm{eff}}$, which is given by%
\begin{equation}
K_{a}^{\mathrm{eff}}=K_{a}-\frac{2\Omega^{2}\omega_{b}}{\omega_{b}^{2}%
+\gamma_{b}^{2}}\;,\label{K_a^eff}%
\end{equation}
is the effective Kerr constant. Taking the module squared of Eq.
(\ref{Eq. for B_a}) leads to%
\begin{equation}
\left[  \left(  \Delta_{a}+K_{a}^{\mathrm{eff}}E_{a}\right)  ^{2}+\left(
\gamma_{a}+\gamma_{a3}E_{a}\right)  ^{2}\right]  E_{a}=2\gamma_{a1}\left\vert
b_{\mathrm{p}}\right\vert ^{2}\;,\label{Eq. for E_a}%
\end{equation}
where%
\begin{equation}
E_{a}=\left\vert B_{a}\right\vert ^{2}\;.
\end{equation}
Finding $E_{a}$ by solving Eq. (\ref{Eq. for E_a}) allows calculating $B_{a}$
according to Eq. (\ref{Eq. for B_a}) and $B_{b}$ according to Eq.
(\ref{Theta_b=0}).

\subsection{Onset of Bistability Point}

In general, for any fixed value of the driving amplitude $b_{\mathrm{p}}$, Eq.
(\ref{Eq. for B_a}) can be expressed as a relation between $E_{a}$ and
$\Delta_{a}$. When $b_{\mathrm{p}}$ is sufficiently large the response of the
system becomes bistable, that is $E_{a}$ becomes a multi-valued function of
$\Delta_{a}$ in some range near the resonance frequency. The onset of
bistability point is defined as the point for which%
\begin{align}
\frac{\partial\Delta_{a}}{\partial E_{a}}  &  =0\ ,\\
\frac{\partial^{2}\Delta_{a}}{\partial\left(  E_{a}\right)  ^{2}}  &  =0\ .
\end{align}
Such a point occurs only if the nonlinear damping is sufficiently small
\cite{Yurke_5054}, namely, only when the following condition holds%
\begin{equation}
\left\vert K_{a}^{\mathrm{eff}}\right\vert >\sqrt{3}\gamma_{a3}\ .
\label{|K_a^eff|>}%
\end{equation}
At the onset of bistability point the drive frequency and amplitude are given
by%
\begin{equation}
\left(  \Delta_{a}\right)  _{c}=-\gamma_{a}\frac{K_{a}^{\mathrm{eff}}%
}{\left\vert K_{a}^{\mathrm{eff}}\right\vert }\left[  \frac{4\gamma_{a3}%
|K_{a}^{\mathrm{eff}}|+\sqrt{3}\left(  \left(  K_{a}^{\mathrm{eff}}\right)
^{2}+\gamma_{a3}^{2}\right)  }{\left(  K_{a}^{\mathrm{eff}}\right)
^{2}-3\gamma_{a3}^{2}}\right]  \ ,
\end{equation}%
\begin{equation}
\left(  b_{\mathrm{p}}\right)  _{c}^{2}=\frac{4}{3\sqrt{3}}\frac{\gamma
_{a}^{3}(\left(  K_{a}^{\mathrm{eff}}\right)  ^{2}+\gamma_{a3}^{2})}%
{\gamma_{a1}\left(  \left\vert K_{a}^{\mathrm{eff}}\right\vert -\sqrt{3}%
\gamma_{a3}\right)  ^{3}}\ , \label{b_p,c}%
\end{equation}
and the resonator mode amplitude is%
\begin{equation}
\left(  E_{a}\right)  _{c}=\frac{2\gamma_{a}}{\sqrt{3}\left(  \left\vert
K_{a}^{\mathrm{eff}}\right\vert -\sqrt{3}\gamma_{a3}\right)  }\ .
\label{|B|_a,c}%
\end{equation}

\subsection{Fluctuation}

Fluctuation around the mean field solution are governed by%
\begin{equation}
\frac{\mathrm{d}}{\mathrm{d}t}\left(
\begin{array}
[c]{c}%
c_{a}\\
c_{a}^{\dag}\\
c_{b}\\
c_{b}^{\dag}%
\end{array}
\right)  +W\left(
\begin{array}
[c]{c}%
c_{a}\\
c_{a}^{\dag}\\
c_{b}\\
c_{b}^{\dag}%
\end{array}
\right)  =\left(
\begin{array}
[c]{c}%
F_{a}\\
F_{a}^{\dag}\\
F_{b}\\
F_{b}^{\dag}%
\end{array}
\right)  \;, \label{dc/dt}%
\end{equation}
where the matrix $W$ is given by%
\begin{equation}
W=\left(
\begin{array}
[c]{cccc}%
W_{1} & W_{2} & W_{3} & W_{4}\\
W_{2}^{\ast} & W_{1}^{\ast} & W_{4}^{\ast} & W_{3}^{\ast}\\
W_{5} & W_{6} & W_{7} & W_{8}\\
W_{6}^{\ast} & W_{5}^{\ast} & W_{8}^{\ast} & W_{7}^{\ast}%
\end{array}
\right)  \;. \label{W}%
\end{equation}
The mean field solution is assumed to be locally stable, that is, it is assume
that all eigenvalues of $W$ have a positive real part.

We calculate below the statistical properties of the noise operators $F_{a}$
and $F_{b}$. Let $a(\omega)$ be an annihilation operator for an incoming bath
mode. In thermal equilibrium the following holds
\begin{align}
\left\langle a\left(  \omega\right)  \right\rangle  &  =0\ ,\\
\left\langle a^{\dagger}\left(  \omega\right)  a\left(  \omega^{\prime
}\right)  \right\rangle  &  =n_{\omega}\delta\left(  \omega-\omega^{\prime
}\right)  \ ,\\
\left\langle a\left(  \omega^{\prime}\right)  a^{\dagger}\left(
\omega\right)  \right\rangle  &  =\left(  n_{\omega}+1\right)  \delta\left(
\omega-\omega^{\prime}\right)  \ ,\\
\left\langle a\left(  \omega\right)  a\left(  \omega^{\prime}\right)
\right\rangle  &  =0\ ,
\end{align}
where
\begin{equation}
n_{\omega}=\frac{1}{e^{\beta\hbar\omega}-1}\ ,
\end{equation}
$\beta=1/k_{\mathrm{B}}T$, $k_{\mathrm{B}}$ is Boltzmann's constant and $T$ is
the absolute temperature. Using these expressions together with Eqs.
(\ref{a^in}), (\ref{a_a1^in rot}), (\ref{a_a2^in rot}), (\ref{a_a3^in rot}),
(\ref{A_a rot}), (\ref{F_a}) and (\ref{F_b}) yields the following relations%
\begin{equation}
\left\langle F_{a}\left(  \omega\right)  \right\rangle =\left\langle
F_{a}^{\dagger}\left(  \omega\right)  \right\rangle =\left\langle F_{b}\left(
\omega\right)  \right\rangle =\left\langle F_{b}^{\dagger}\left(
\omega\right)  \right\rangle =0\ , \label{<F>=<F+>=0}%
\end{equation}%
\begin{align}
\left\langle F_{a}\left(  \omega\right)  F_{a}\left(  \omega^{\prime}\right)
\right\rangle  &  =\left\langle F_{a}^{\dagger}\left(  \omega\right)
F_{a}^{\dagger}\left(  \omega^{\prime}\right)  \right\rangle \nonumber\\
&  =\left\langle F_{b}\left(  \omega\right)  F_{b}\left(  \omega^{\prime
}\right)  \right\rangle =\left\langle F_{b}^{\dagger}\left(  \omega\right)
F_{b}^{\dagger}\left(  \omega^{\prime}\right)  \right\rangle =0\ ,\nonumber\\
&  \label{<FF>=0}%
\end{align}%
\begin{equation}
\left\langle F_{a}\left(  \omega\right)  F_{a}^{\dagger}\left(  \omega
^{\prime}\right)  \right\rangle =2\Gamma_{a}\delta\left(  \omega
-\omega^{\prime}\right)  n_{\omega_{a}}\ , \label{<F_aF_a+>}%
\end{equation}%
\begin{equation}
\left\langle F_{a}^{\dagger}\left(  \omega\right)  F_{a}\left(  \omega
^{\prime}\right)  \right\rangle =2\Gamma_{a}\delta\left(  \omega
-\omega^{\prime}\right)  \left(  n_{\omega_{a}}+1\right)  \ .
\label{<F_a+F_a>}%
\end{equation}%
\begin{equation}
\left\langle F_{b}\left(  \omega\right)  F_{b}^{\dagger}\left(  \omega
^{\prime}\right)  \right\rangle =2\gamma_{b}\delta\left(  \omega
-\omega^{\prime}\right)  n_{\omega_{b}}\ , \label{<F_bF_b+>}%
\end{equation}
and%
\begin{equation}
\left\langle F_{b}^{\dagger}\left(  \omega\right)  F_{b}\left(  \omega
^{\prime}\right)  \right\rangle =2\gamma_{b}\delta\left(  \omega
-\omega^{\prime}\right)  \left(  n_{\omega_{b}}+1\right)  \ ,
\label{<F_b+F_b>}%
\end{equation}
where%
\begin{equation}
\Gamma_{a}=\gamma_{a}+2\gamma_{a3}E_{a}\;. \label{Gamma_a}%
\end{equation}

Is is important to note that the linearization approach is valid only when the
fluctuations around the mean field solution are small. Unavoidably, however,
very close to the region where the system becomes unstable the fluctuations
become appreciable, and consequently the linearization approximation breaks down.

\subsection{Transforming into Fourier space}

In general, the Fourier transform of a time dependent operator $O\left(
t\right)  $ is denoted as $O\left(  \omega\right)  $%
\begin{equation}
O\left(  t\right)  =\frac{1}{\sqrt{2\pi}}\int\limits_{-\infty}^{\infty
}\mathrm{d}\omega\;O\left(  \omega\right)  e^{-i\omega t}\;.
\end{equation}
Applying the Fourier transform to Eq. (\ref{dc/dt}) yields%
\begin{equation}
W_{aa}\left(
\begin{array}
[c]{c}%
c_{a}\left(  \omega\right)  \\
c_{a}^{\dag}\left(  -\omega\right)
\end{array}
\right)  +W_{ab}\left(
\begin{array}
[c]{c}%
c_{b}\left(  \omega\right)  \\
c_{b}^{\dag}\left(  -\omega\right)
\end{array}
\right)  =\left(
\begin{array}
[c]{c}%
F_{a}\left(  \omega\right)  \\
F_{a}^{\dag}\left(  -\omega\right)
\end{array}
\right)  \;,
\end{equation}%
\begin{equation}
W_{ba}\left(
\begin{array}
[c]{c}%
c_{a}\left(  \omega\right)  \\
c_{a}^{\dag}\left(  -\omega\right)
\end{array}
\right)  +W_{bb}\left(
\begin{array}
[c]{c}%
c_{b}\left(  \omega\right)  \\
c_{b}^{\dag}\left(  -\omega\right)
\end{array}
\right)  =\left(
\begin{array}
[c]{c}%
F_{b}\left(  \omega\right)  \\
F_{b}^{\dag}\left(  -\omega\right)
\end{array}
\right)  \;,
\end{equation}
where%
\begin{align}
W_{aa} &  =\left(
\begin{array}
[c]{cc}%
W_{1}-i\omega & W_{2}\\
W_{2}^{\ast} & W_{1}^{\ast}-i\omega
\end{array}
\right)  \;,\label{W_aa}\\
W_{ab} &  =\left(
\begin{array}
[c]{cc}%
W_{3} & W_{4}\\
W_{4}^{\ast} & W_{3}^{\ast}%
\end{array}
\right)  \;,\label{W_ab}\\
W_{ba} &  =\left(
\begin{array}
[c]{cc}%
W_{5} & W_{6}\\
W_{6}^{\ast} & W_{5}^{\ast}%
\end{array}
\right)  \;,\label{W_ba}\\
W_{bb} &  =\left(
\begin{array}
[c]{cc}%
W_{7}-i\omega & W_{8}\\
W_{8}^{\ast} & W_{7}^{\ast}-i\omega
\end{array}
\right)  \;.\label{W_bb}%
\end{align}
Multiplying the first equation by $W_{aa}^{-1}$ and the second one by
$W_{bb}^{-1}$ leads to%
\begin{equation}
\left(
\begin{array}
[c]{c}%
c_{a}\left(  \omega\right)  \\
c_{a}^{\dag}\left(  -\omega\right)
\end{array}
\right)  +W_{aa}^{-1}W_{ab}\left(
\begin{array}
[c]{c}%
c_{b}\left(  \omega\right)  \\
c_{b}^{\dag}\left(  -\omega\right)
\end{array}
\right)  =W_{aa}^{-1}\left(
\begin{array}
[c]{c}%
F_{a}\left(  \omega\right)  \\
F_{a}^{\dag}\left(  -\omega\right)
\end{array}
\right)  \;,
\end{equation}%
\begin{equation}
\left(
\begin{array}
[c]{c}%
c_{b}\left(  \omega\right)  \\
c_{b}^{\dag}\left(  -\omega\right)
\end{array}
\right)  +W_{bb}^{-1}W_{ba}\left(
\begin{array}
[c]{c}%
c_{a}\left(  \omega\right)  \\
c_{a}^{\dag}\left(  -\omega\right)
\end{array}
\right)  =W_{bb}^{-1}\left(
\begin{array}
[c]{c}%
F_{b}\left(  \omega\right)  \\
F_{b}^{\dag}\left(  -\omega\right)
\end{array}
\right)  \;,
\end{equation}
or%
\begin{align}
&  \left(
\begin{array}
[c]{c}%
c_{a}\left(  \omega\right)  \\
c_{a}^{\dag}\left(  -\omega\right)
\end{array}
\right)  =\chi_{aa}\left(
\begin{array}
[c]{c}%
F_{a}\left(  \omega\right)  \\
F_{a}^{\dag}\left(  -\omega\right)
\end{array}
\right)  +\chi_{ab}\left(
\begin{array}
[c]{c}%
F_{b}\left(  \omega\right)  \\
F_{b}^{\dag}\left(  -\omega\right)
\end{array}
\right)  \;,\nonumber\\
& \label{c_a(omega)}%
\end{align}%
\begin{align}
&  \left(
\begin{array}
[c]{c}%
c_{b}\left(  \omega\right)  \\
c_{b}^{\dag}\left(  -\omega\right)
\end{array}
\right)  =\chi_{ba}\left(
\begin{array}
[c]{c}%
F_{a}\left(  \omega\right)  \\
F_{a}^{\dag}\left(  -\omega\right)
\end{array}
\right)  +\chi_{bb}\left(
\begin{array}
[c]{c}%
F_{b}\left(  \omega\right)  \\
F_{b}^{\dag}\left(  -\omega\right)
\end{array}
\right)  \;.\nonumber\\
& \label{c_b(omega)}%
\end{align}
where%
\begin{subequations}
\begin{align}
\chi_{aa} &  =\left(  W_{aa}-W_{ab}W_{bb}^{-1}W_{ba}\right)  ^{-1}%
\;,\label{chi_aa}\\
\chi_{ab} &  =\left(  W_{ba}-W_{bb}W_{ab}^{-1}W_{aa}\right)  ^{-1}%
\;,\label{chi_ab}\\
\chi_{ba} &  =\left(  W_{ab}-W_{aa}W_{ba}^{-1}W_{bb}\right)  ^{-1}%
\;,\label{chi_ba}\\
\chi_{bb} &  =\left(  W_{bb}-W_{ba}W_{aa}^{-1}W_{ab}\right)  ^{-1}%
\;.\label{chi_bb}%
\end{align}

The inverse matrices $W_{aa}^{-1}$ and $W_{bb}^{-1}$ can be expressed as%
\end{subequations}
\begin{align}
W_{aa}^{-1}  &  =\frac{\left(
\begin{array}
[c]{cc}%
W_{1}^{\ast}-i\omega & -W_{2}\\
-W_{2}^{\ast} & W_{1}-i\omega
\end{array}
\right)  }{\left(  W_{1}-i\omega\right)  \left(  W_{1}^{\ast}-i\omega\right)
-\left\vert W_{2}\right\vert ^{2}}\nonumber\\
&  =\frac{\left(
\begin{array}
[c]{cc}%
W_{1}^{\ast}-i\omega & -W_{2}\\
-W_{2}^{\ast} & W_{1}-i\omega
\end{array}
\right)  }{\left(  \lambda_{a1}-i\omega\right)  \left(  \lambda_{a2}%
-i\omega\right)  }\;,\nonumber\\
&
\end{align}%
\begin{align}
W_{bb}^{-1}  &  =\frac{\left(
\begin{array}
[c]{cc}%
W_{7}^{\ast}-i\omega & -W_{8}\\
-W_{8}^{\ast} & W_{7}-i\omega
\end{array}
\right)  }{\left(  W_{7}-i\omega\right)  \left(  W_{7}^{\ast}-i\omega\right)
-\left\vert W_{8}\right\vert ^{2}}\nonumber\\
&  =\frac{\left(
\begin{array}
[c]{cc}%
W_{7}^{\ast}-i\omega & -W_{8}\\
-W_{8}^{\ast} & W_{7}-i\omega
\end{array}
\right)  }{\left(  \lambda_{b1}-i\omega\right)  \left(  \lambda_{b2}%
-i\omega\right)  }\;,\nonumber\\
&
\end{align}
where we have introduced the eigenvalues%
\begin{subequations}
\begin{align}
\lambda_{a1}+\lambda_{a2}  &  =W_{1}+W_{1}^{\ast}\;,\\
\lambda_{a1}\lambda_{a2}  &  =\left\vert W_{1}\right\vert ^{2}-\left\vert
W_{2}\right\vert ^{2}\;,
\end{align}
and%
\end{subequations}
\begin{subequations}
\begin{align}
\lambda_{b1}+\lambda_{b2}  &  =W_{7}+W_{7}^{\ast}\;,\\
\lambda_{b1}\lambda_{b2}  &  =\left\vert W_{7}\right\vert ^{2}-\left\vert
W_{8}\right\vert ^{2}\;.
\end{align}

\subsection{Omega-Symmetric Matrix}

Let $W\left(  \omega\right)  $ be a 2X2 matrix, which depends on the real
parameter $\omega$. The matrix $W\left(  \omega\right)  $ is said to be
omega-symmetric if it can be written as%
\end{subequations}
\begin{equation}
W\left(  \omega\right)  =\left(
\begin{array}
[c]{cc}%
a\left(  \omega\right)  & b\left(  \omega\right) \\
b^{\ast}\left(  -\omega\right)  & a^{\ast}\left(  -\omega\right)
\end{array}
\right)  \;,
\end{equation}
where $a\left(  \omega\right)  $ and $b\left(  \omega\right)  $ are arbitrary
smooth functions of $\omega$. Is is straightforward to show that if $W$ is
omega-symmetric then $W^{-1}$, $W^{t}$ (transpose of $W$) and $W^{\dag}$ are
all omega-symmetric as well. Moreover, if $W_{1}$ and $W_{2}$ are both
omega-symmetric then $W_{1}W_{2}$ is also omega-symmetric. Thus, it is easy to
show that the susceptibility matrixes $\chi_{aa}$, $\chi_{ab}$, $\chi_{ba}$
and $\chi_{bb}$ are all omega-symmetric.

\subsection{The case where $\Omega$ is small and $K_{a}=\gamma_{a3}=0$}

To lowest order in $\Omega$ one has%
\begin{align}
\chi_{aa}  &  =\left(  1-W_{aa}^{-1}W_{ab}W_{bb}^{-1}W_{ba}\right)
^{-1}W_{aa}^{-1}\nonumber\\
&  \simeq\left(  1+W_{aa}^{-1}W_{ab}W_{bb}^{-1}W_{ba}\right)  W_{aa}%
^{-1}\;,\nonumber\\
&
\end{align}%
\begin{equation}
\chi_{ab}\simeq-W_{aa}^{-1}W_{ab}W_{bb}^{-1}\;,
\end{equation}%
\begin{equation}
\chi_{ba}\simeq-W_{bb}^{-1}W_{ba}W_{aa}^{-1}\;,
\end{equation}
and%
\begin{align}
\chi_{bb}  &  =\left(  1-W_{bb}^{-1}W_{ba}W_{aa}^{-1}W_{ab}\right)
^{-1}W_{bb}^{-1}\nonumber\\
&  \simeq\left(  1+W_{bb}^{-1}W_{ba}W_{aa}^{-1}W_{ab}\right)  W_{bb}%
^{-1}\;.\nonumber\\
&
\end{align}

Taking $K_{a}=\gamma_{a3}=0$ one has%
\begin{equation}
W_{aa}^{-1}=\left(
\begin{array}
[c]{cc}%
\frac{1}{\lambda_{a1}-i\omega} & 0\\
0 & \frac{1}{\lambda_{a2}-i\omega}%
\end{array}
\right)  \;.
\end{equation}
Similarly $W_{bb}^{-1}$ can be expressed as%
\begin{equation}
W_{bb}^{-1}=\left(
\begin{array}
[c]{cc}%
\frac{1}{\lambda_{b1}-i\omega} & 0\\
0 & \frac{1}{\lambda_{b2}-i\omega}%
\end{array}
\right)  \;.
\end{equation}
Using these relations one finds that%
\begin{align}
\chi_{aa}  &  =\left(
\begin{array}
[c]{cc}%
\frac{1}{\lambda_{a1}-i\omega} & 0\\
0 & \frac{1}{\lambda_{a2}-i\omega}%
\end{array}
\right) \nonumber\\
&  +\frac{\Omega^{2}\left(
\begin{array}
[c]{cc}%
\frac{E_{a}\left(  \lambda_{b1}-\lambda_{b2}\right)  }{\left(  \lambda
_{a1}-i\omega\right)  ^{2}} & \frac{B_{a}^{2}\left(  \lambda_{b1}-\lambda
_{b2}\right)  }{\left(  \lambda_{a1}-i\omega\right)  \left(  \lambda
_{a2}-i\omega\right)  }\\
-\frac{\left(  B_{a}^{\ast}\right)  ^{2}\left(  \lambda_{b1}-\lambda
_{b2}\right)  }{\left(  \lambda_{a1}-i\omega\right)  \left(  \lambda
_{a2}-i\omega\right)  } & -\frac{E_{a}\left(  \lambda_{b1}-\lambda
_{b2}\right)  }{\left(  \lambda_{a2}-i\omega\right)  ^{2}}%
\end{array}
\right)  }{\left(  \lambda_{b1}-i\omega\right)  \left(  \lambda_{b2}%
-i\omega\right)  }\;,\nonumber\\
&  \label{chi_aa app}%
\end{align}%
\begin{equation}
\chi_{ab}=-\Omega\left(
\begin{array}
[c]{cc}%
\frac{iB_{a}}{\left(  \lambda_{a1}-i\omega\right)  \left(  \lambda
_{b1}-i\omega\right)  } & \frac{iB_{a}}{\left(  \lambda_{a1}-i\omega\right)
\left(  \lambda_{b2}-i\omega\right)  }\\
-\frac{iB_{a}^{\ast}}{\left(  \lambda_{a2}-i\omega\right)  \left(
\lambda_{b1}-i\omega\right)  } & -\frac{iB_{a}^{\ast}}{\left(  \lambda
_{a2}-i\omega\right)  \left(  \lambda_{b2}-i\omega\right)  }%
\end{array}
\right)  \;, \label{chi_ab app}%
\end{equation}%
\begin{equation}
\chi_{ba}=-\Omega\left(
\begin{array}
[c]{cc}%
\frac{iB_{a}^{\ast}}{\left(  \lambda_{a1}-i\omega\right)  \left(  \lambda
_{b1}-i\omega\right)  } & \frac{iB_{a}}{\left(  \lambda_{a2}-i\omega\right)
\left(  \lambda_{b1}-i\omega\right)  }\\
-\frac{iB_{a}^{\ast}}{\left(  \lambda_{a1}-i\omega\right)  \left(
\lambda_{b2}-i\omega\right)  } & -\frac{iB_{a}}{\left(  \lambda_{a2}%
-i\omega\right)  \left(  \lambda_{b2}-i\omega\right)  }%
\end{array}
\right)  \; \label{chi_ba app}%
\end{equation}
and%
\begin{align}
\chi_{bb}  &  =\left(
\begin{array}
[c]{cc}%
\frac{1}{\lambda_{b1}-i\omega} & 0\\
0 & \frac{1}{\lambda_{b2}-i\omega}%
\end{array}
\right) \nonumber\\
&  +\frac{\Omega^{2}E_{a}\left(
\begin{array}
[c]{cc}%
\frac{\left(  \lambda_{a1}-\lambda_{a2}\right)  }{\left(  \lambda_{b1}%
-i\omega\right)  ^{2}} & \frac{\left(  \lambda_{a1}-\lambda_{a2}\right)
}{\left(  \lambda_{b1}-i\omega\right)  \left(  \lambda_{b2}-i\omega\right)
}\\
-\frac{\left(  \lambda_{a1}-\lambda_{a2}\right)  }{\left(  \lambda
_{b1}-i\omega\right)  \left(  \lambda_{b2}-i\omega\right)  } & -\frac{\left(
\lambda_{a1}-\lambda_{a2}\right)  }{\left(  \lambda_{b2}-i\omega\right)  ^{2}}%
\end{array}
\right)  }{\left(  \lambda_{a1}-i\omega\right)  \left(  \lambda_{a2}%
-i\omega\right)  }\;.\nonumber\\
&  \label{chi_bb app}%
\end{align}

To determine the stability of the mean field solutions the eigenvalues of $W$
are calculated below for the present case to lowest nonvanishing order in
$\Omega$. The matrix $W$ can be expressed as%
\begin{equation}
W=\left(
\begin{array}
[c]{cccc}%
\lambda_{a1} & 0 & 0 & 0\\
0 & \lambda_{a2} & 0 & 0\\
0 & 0 & \lambda_{b1} & 0\\
0 & 0 & 0 & \lambda_{b2}%
\end{array}
\right)  +\Omega V\;.
\end{equation}
where%
\[
V=\left(
\begin{array}
[c]{cccc}%
0 & 0 & iB_{a} & iB_{a}\\
0 & 0 & -iB_{a}^{\ast} & -iB_{a}^{\ast}\\
iB_{a}^{\ast} & iB_{a} & 0 & 0\\
-iB_{a}^{\ast} & -iB_{a} & 0 & 0
\end{array}
\right)  \;.
\]
The two eigenvalues of interest for what follows are $\tilde{\lambda}_{b1}$
and $\tilde{\lambda}_{b2}$, which approach the values $\lambda_{b1}$ and
$\lambda_{b2}$ respectively in the limit $\Omega\rightarrow0$. These
eigenvalues are calculated up to second order in $\Omega$ using perturbation
theory (note that $W$ is not necessarily Hermitian)%
\begin{subequations}
\begin{align}
\tilde{\lambda}_{b1}  &  =\lambda_{b1}+\Omega^{2}E_{a}\left(  -\frac
{1}{\lambda_{b1}-\lambda_{a1}}+\frac{1}{\lambda_{b1}-\lambda_{a2}}\right)
\;,\\
\tilde{\lambda}_{b2}  &  =\lambda_{b2}+\Omega^{2}E_{a}\left(  \frac{1}%
{\lambda_{b2}-\lambda_{a1}}-\frac{1}{\lambda_{b2}-\lambda_{a2}}\right)  \;.
\end{align}
Thus by using the relations%
\end{subequations}
\begin{subequations}
\begin{align}
\lambda_{a1}  &  =\lambda_{a2}^{\ast}=i\Delta_{a}^{\mathrm{eff}}+\gamma
_{a}\;,\\
\lambda_{b1}  &  =\lambda_{b1}^{\ast}=i\omega_{b}+\gamma_{b}\;,
\end{align}
the notation%
\end{subequations}
\begin{subequations}
\begin{align}
d  &  =\frac{\Delta_{a}^{\mathrm{eff}}}{\omega_{b}}\ ,\\
g  &  =\frac{\gamma_{a}}{\omega_{b}}\ ,
\end{align}
and by assuming also that $\gamma_{b}\ll\omega_{b}$ one finds that%
\end{subequations}
\begin{align}
\tilde{\lambda}_{b1}  &  =i\omega_{b}\left(  1+\frac{2\Omega^{2}E_{a}}%
{\omega_{b}^{2}}\frac{2d\left(  1-d^{2}-g^{2}\right)  }{\left[  \left(
d+1\right)  ^{2}+g^{2}\right]  \left[  \left(  d-1\right)  ^{2}+g^{2}\right]
}\right) \nonumber\\
&  +\gamma_{b}\left(  1+\frac{2\Omega^{2}E_{a}}{\gamma_{a}\gamma_{b}}%
\frac{4dg^{2}}{\left[  \left(  1+d\right)  ^{2}+g^{2}\right]  \left[  \left(
1-d\right)  ^{2}+g^{2}\right]  }\right)  \;,\nonumber\\
&
\end{align}
and $\tilde{\lambda}_{b2}=\tilde{\lambda}_{b1}^{\ast}$.

\begin{figure}
[ptb]
\begin{center}
\includegraphics[
height=2.8738in,
width=3.2396in
]%
{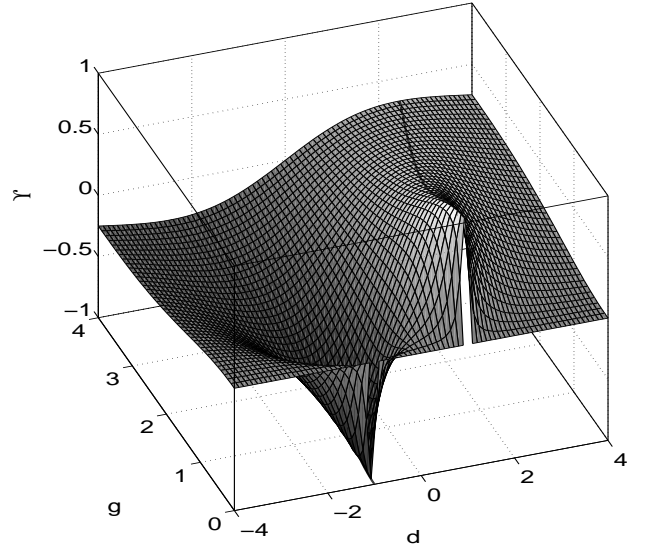}%
\caption{The function $\Upsilon\left(  d,g\right)  $.}%
\label{Ups}%
\end{center}
\end{figure}

For the present case ($K_{a}=\gamma_{a3}=0$) one finds using Eqs.
(\ref{K_a^eff}) and (\ref{|B|_a,c}) that $\left(  E_{a}\right)  _{c}$ (the
value of $E_{a}$ at the onset of bistability) is given by%
\begin{equation}
\left(  E_{a}\right)  _{c}=\frac{\gamma_{a}\omega_{b}}{\sqrt{3}\Omega^{2}}\;.
\label{(E_a)_c}%
\end{equation}
In terms of $\left(  E_{a}\right)  _{c}$ the real part of $\tilde{\lambda
}_{b1}$ can be expressed as%
\begin{equation}
\frac{\operatorname{Re}\left(  \tilde{\lambda}_{b1}\right)  }{\gamma_{b}%
}=1+\frac{2E_{a}}{\sqrt{3}\left(  E_{a}\right)  _{c}}\frac{\omega_{b}}%
{\gamma_{b}}\Upsilon\left(  d,g\right)  \;,
\end{equation}
where the function $\Upsilon\left(  d,g\right)  $, which is plotted in Fig.
\ref{Ups}, is given by%

\begin{align}
\Upsilon\left(  d,g\right)   &  =\frac{4dg^{2}}{\left[  \left(  1+d\right)
^{2}+g^{2}\right]  \left[  \left(  1-d\right)  ^{2}+g^{2}\right]  }\nonumber\\
&  =\frac{4g^{2}d}{4g^{2}+\left(  g^{2}-1+d^{2}\right)  ^{2}}\;.\nonumber\\
&
\end{align}
For any given value of $g$ the function $\Upsilon$ obtains a maxima at
$d=d_{0}$ and a minima at $d=-d_{0}$, where%
\begin{equation}
d_{0}=\frac{1}{3}\sqrt{3-3g^{2}+6\sqrt{g^{4}+g^{2}+1}}\;.
\end{equation}
The mean field solution is stable provided that $\operatorname{Re}\left(
\tilde{\lambda}_{b1}\right)  >0$. Hopf bifurcation occurs when
$\operatorname{Re}\left(  \tilde{\lambda}_{b1}\right)  $ vanishes.%

\section{Integrated Spectral Density}

In general consider an operator $c\left(  \omega\right)  $ that can be
expressed in terms of a noise operator $F\left(  \omega\right)  $ and a
susceptibility matrix $\chi\left(  \omega\right)  $ as [similarly to Eqs.
(\ref{c_a(omega)})\ and (\ref{c_b(omega)})]%
\begin{equation}
\left(
\begin{array}
[c]{c}%
c\left(  \omega\right) \\
c^{\dag}\left(  -\omega\right)
\end{array}
\right)  =\chi\left(  \omega\right)  \left(
\begin{array}
[c]{c}%
F\left(  \omega\right) \\
F^{\dag}\left(  -\omega\right)
\end{array}
\right)  \;,
\end{equation}
where $F\left(  \omega\right)  $ satisfy [similarly to Eqs. (\ref{<F>=<F+>=0}%
), (\ref{<FF>=0}), (\ref{<F_aF_a+>}), (\ref{<F_a+F_a>}), (\ref{<F_bF_b+>}) and
(\ref{<F_b+F_b>})]%
\begin{equation}
\left\langle F\left(  \omega\right)  \right\rangle =\left\langle F^{\dagger
}\left(  \omega\right)  \right\rangle =0\ ,
\end{equation}%
\begin{equation}
\left\langle F\left(  \omega\right)  F\left(  \omega^{\prime}\right)
\right\rangle =\left\langle F^{\dagger}\left(  \omega\right)  F^{\dagger
}\left(  \omega^{\prime}\right)  \right\rangle =0\ ,
\end{equation}%
\begin{equation}
\left\langle F\left(  \omega\right)  F^{\dagger}\left(  \omega^{\prime
}\right)  \right\rangle =2\Gamma\delta\left(  \omega-\omega^{\prime}\right)
n_{\omega_{0}}\ ,
\end{equation}
and%
\begin{equation}
\left\langle F^{\dagger}\left(  \omega\right)  F\left(  \omega^{\prime
}\right)  \right\rangle =2\Gamma\delta\left(  \omega-\omega^{\prime}\right)
\left(  n_{\omega_{0}}+1\right)  \ .
\end{equation}

The homodyne detection observable $X\left(  \omega\right)  $ is defined by%
\begin{equation}
X\left(  \omega\right)  =e^{i\phi_{\mathrm{LO}}}c\left(  \omega\right)
+e^{-i\phi_{\mathrm{LO}}}c^{\dag}\left(  -\omega\right)  \ .
\end{equation}
The frequency auto-correlation function of $X$ is related to the spectral
density $P_{X}\left(  \omega\right)  $ by%
\begin{equation}
\left\langle X^{\dag}\left(  \omega^{\prime}\right)  X\left(  \omega\right)
\right\rangle =P_{X}\left(  \omega\right)  \delta\left(  \omega-\omega
^{\prime}\right)  \ .
\end{equation}
Assuming that $\chi\left(  \omega\right)  $ is omega-symmetric, it can be
expressed as%
\begin{equation}
\chi\left(  \omega\right)  =\left(
\begin{array}
[c]{cc}%
a\left(  \omega\right)   & b\left(  \omega\right)  \\
b^{\ast}\left(  -\omega\right)   & a^{\ast}\left(  -\omega\right)
\end{array}
\right)  \;,
\end{equation}
where $a\left(  \omega\right)  $ and $b\left(  \omega\right)  $ are arbitrary
functions of $\omega$. Thus, by calculating the term $\left\langle X^{\dag
}\left(  \omega^{\prime}\right)  X\left(  \omega\right)  \right\rangle $ one
finds that%
\begin{align}
\frac{P_{X}\left(  \omega\right)  }{2\Gamma} &  =M_{+}\left(  \omega\right)
\coth\frac{\beta\hbar\omega_{0}}{2}+M_{-}\left(  \omega\right)  \ .\nonumber\\
& \label{P(omega) M+ M-}%
\end{align}
where%
\begin{align}
M_{+}\left(  \omega\right)   &  =\frac{\left\vert a\left(  -\omega\right)
\right\vert ^{2}+\left\vert b\left(  \omega\right)  \right\vert ^{2}%
+\left\vert a\left(  \omega\right)  \right\vert ^{2}+\left\vert b\left(
-\omega\right)  \right\vert ^{2}}{2}\nonumber\\
&  +\operatorname{Re}\left[  e^{2i\phi_{\mathrm{LO}}}\left(  a\left(
-\omega\right)  b\left(  \omega\right)  +a\left(  \omega\right)  b\left(
-\omega\right)  \right)  \right]  \;,\nonumber\\
& \label{M+}%
\end{align}
and%
\begin{align}
M_{-}\left(  \omega\right)   &  =\frac{-\left\vert a\left(  -\omega\right)
\right\vert ^{2}-\left\vert b\left(  \omega\right)  \right\vert ^{2}%
+\left\vert a\left(  \omega\right)  \right\vert ^{2}+\left\vert b\left(
-\omega\right)  \right\vert ^{2}}{2}\nonumber\\
&  +\operatorname{Re}\left[  e^{2i\phi_{\mathrm{LO}}}\left(  -a\left(
-\omega\right)  b\left(  \omega\right)  +a\left(  \omega\right)  b\left(
-\omega\right)  \right)  \right]  \;.\nonumber\\
& \label{M-}%
\end{align}
The integrated spectral density (ISD) is thus given by%
\begin{equation}
\int\limits_{-\infty}^{\infty}\mathrm{d}\omega\;P_{X}\left(  \omega\right)
=2\Gamma V\coth\frac{\beta\hbar\omega_{0}}{2}\ ,\label{int P}%
\end{equation}
where%
\begin{align}
V &  =\int\limits_{-\infty}^{\infty}\mathrm{d}\omega\;M_{+}\left(
\omega\right)  \nonumber\\
&  =\int\limits_{-\infty}^{\infty}\mathrm{d}\omega\;\left[  \left\vert
a\left(  \omega\right)  \right\vert ^{2}+\left\vert b\left(  -\omega\right)
\right\vert ^{2}+2\operatorname{Re}\left(  e^{2i\phi_{\mathrm{LO}}}a\left(
\omega\right)  b\left(  -\omega\right)  \right)  \right]  \ .\nonumber\\
& \label{V}%
\end{align}

\section{ISD of $X_{b}$}

We calculate below the ISD of the homodyne observable $X_{b}\left(
\omega\right)  $, which is given by%
\begin{equation}
X_{b}\left(  \omega\right)  =e^{i\phi_{\mathrm{LO}}}c_{b}\left(
\omega\right)  +e^{-i\phi_{\mathrm{LO}}}c_{b}^{\dag}\left(  -\omega\right)
\ ,
\end{equation}
for the case where $\Omega$ is small and $K_{a}=\gamma_{a3}=0$. As can be seen
from Eq. (\ref{c_b(omega)}), it has two contributions due to the two
uncorrelated noise terms $F_{b}\left(  \omega\right)  $ and $F_{a}\left(
\omega\right)  $. The calculation of both contributions according to Eq.
(\ref{int P}) is involved with evaluation of some integrals, which can be
performed using the residue theorem. To further simplify the final result,
which is given by
\begin{align}
&  \frac{1}{2\pi}\int\limits_{-\infty}^{\infty}\mathrm{d}\omega\;P_{X_{b}%
}\left(  \omega\right) \nonumber\\
&  =\left(  1-\frac{\Omega^{2}E_{a}}{\gamma_{a}\gamma_{b}}\Upsilon\left(
d,g\right)  \right)  \coth\frac{\beta\hbar\omega_{b}}{2}\nonumber\\
&  +\frac{\Omega^{2}E_{a}}{\gamma_{a}\gamma_{b}}\frac{2g^{2}}{\left(
1-d\right)  ^{2}+g^{2}}\coth\frac{\beta\hbar\omega_{a}}{2}\ ,\nonumber\\
&  \label{ISD gb ss ga}%
\end{align}
the case where resonator $b$ has high quality factor is assume. For this case,
which is experimentally common, the following is assumed to hold $\gamma
_{b}\ll\omega_{b}$ and $\gamma_{b}\ll\gamma_{a}$. As can be seen from Eq.
(\ref{ISD gb ss ga}), for finite driven amplitude $E_{a}$ the ISD of $X_{b}$
can deviate from the equilibrium value of $\coth\left(  \beta\hbar\omega
_{b}/2\right)  $.

\section{Decoherence}

The Hamiltonian of the system (\ref{Hamiltonian}) is formally a function of
$A_{b}$ and $A_{b}^{\dagger}$, that is $\mathcal{H}=\mathcal{H}\left(
A_{b},A_{b}^{\dag}\right)  $. Consider resonator $b$ in a superposition of two
coherent states $\left\vert \alpha_{1}\right\rangle $ and $\left\vert
\alpha_{2}\right\rangle $. In therm of the operator $\mathcal{V}$, which is
given by
\begin{equation}
\mathcal{V}=\mathcal{H}\left(  \alpha_{2},\alpha_{2}^{\ast}\right)
-\mathcal{H}\left(  \alpha_{1},\alpha_{1}^{\ast}\right)  \;,
\end{equation}
the decoherence rate $1/\tau_{\varphi}$ can be expressed as
\cite{Levinson_299}%
\begin{equation}
\frac{1}{\tau_{\varphi}}=\frac{1}{\hbar^{2}}\int_{-\infty}^{\infty}%
\mathrm{d}\omega\;\left\langle \mathcal{\tilde{V}}\left(  0\right)
\mathcal{\tilde{V}}\left(  \omega\right)  \right\rangle \;,
\end{equation}
where%
\begin{equation}
\mathcal{\tilde{V}}\left(  t\right)  =\mathcal{V}\left(  t\right)
-\left\langle \mathcal{V}\left(  t\right)  \right\rangle \;,
\end{equation}
and where $\mathcal{\tilde{V}}\left(  \omega\right)  $ is the Fourier
transform of $\mathcal{\tilde{V}}\left(  t\right)  $%
\begin{equation}
\mathcal{\tilde{V}}\left(  t\right)  =\frac{1}{\sqrt{2\pi}}\int_{-\infty
}^{\infty}\mathrm{d}\omega\;\mathcal{\tilde{V}}\left(  \omega\right)
e^{-i\omega t}\;.
\end{equation}

Using Eqs. (\ref{F_b}) and (\ref{c_a(omega)}) together with the notation%
\begin{equation}
\delta_{\alpha}=\alpha_{2}-\alpha_{1}=\left\vert \delta_{\alpha}\right\vert
e^{i\theta}\;,
\end{equation}
one finds to lowest order that%
\begin{equation}
\frac{\mathcal{\tilde{V}}\left(  \omega\right)  }{\hbar\left\vert
\delta_{\alpha}\right\vert }=U_{a}F_{a}\left(  \omega\right)  +U_{a}^{\ast
}F_{a}^{\dag}\left(  -\omega\right)  +U_{b}F_{b}\left(  \omega\right)
+U_{b}^{\ast}F_{b}^{\dag}\left(  -\omega\right)  \;,
\end{equation}
where%
\begin{subequations}
\begin{align}
U_{a} &  =2\Omega\cos\theta\left(  B_{a}^{\ast}\left(  \chi_{aa}\right)
_{11}+B_{a}\left(  \chi_{aa}\right)  _{21}\right)  \;,\\
U_{b} &  =2\Omega\cos\theta\left(  B_{a}^{\ast}\left(  \chi_{ab}\right)
_{11}+B_{a}\left(  \chi_{ab}\right)  _{21}\right)  +ie^{-i\theta}\;.
\end{align}
Furthermore, with the help of Eqs. (\ref{<F_aF_a+>}), (\ref{<F_a+F_a>}),
(\ref{<F_bF_b+>}) and (\ref{<F_b+F_b>}) the decoherence rate becomes%
\end{subequations}
\begin{align}
\frac{1}{\tau_{\varphi}} &  =2\left\vert \delta_{\alpha}\right\vert
^{2}\left(  \Gamma_{a}\left\vert U_{a}\right\vert ^{2}\coth\frac{\beta
\hbar\omega_{a}}{2}+\gamma_{b}\left\vert U_{b}\right\vert ^{2}\coth\frac
{\beta\hbar\omega_{b}}{2}\right)  \;.\nonumber\\
& \label{1/tau GC}%
\end{align}
Note that for $\Omega=0$ the decoherence rate reproduces the value given by
Eq. (\ref{1/tau TE}).

For the case where $\Omega$ is small and $K_{a}=\gamma_{a3}=0$ one finds using
Eqs. (\ref{chi_aa app}) and (\ref{chi_ab app}) that%
\begin{equation}
\left\vert U_{a}\right\vert ^{2}=\frac{4\Omega^{2}E_{a}\cos^{2}\theta}%
{\omega_{b}^{2}\left(  d^{2}+g^{2}\right)  }\;,
\end{equation}
and%
\begin{equation}
\left\vert U_{b}\right\vert ^{2}=1+\frac{4\Omega^{2}E_{a}\left[  \cos
a+\cos\left(  2\theta+a\right)  \right]  }{\omega_{b}^{2}\sqrt{1+\left(
\frac{\gamma_{b}}{\omega_{b}}\right)  ^{2}}}\frac{d}{d^{2}+g^{2}}\;.
\end{equation}
where%
\begin{equation}
a=\tan^{-1}\frac{\gamma_{b}}{\omega_{b}}\;.
\end{equation}

In what follows we restrict the discussion to the case where $\theta=0$, for
which the two coherent states $\left\vert \alpha_{1}\right\rangle $ and
$\left\vert \alpha_{2}\right\rangle $ have the same momentum. For this case,
which is the assumed case in some of the published proposals for observation
of quantum superposition in mechanical systems
\cite{Bose_4175,Bose_3204,Buks_174504}, up to first order in $\gamma
_{b}/\omega_{b}$ one has%
\begin{equation}
\left\vert U_{b}\right\vert ^{2}=1+\frac{4\Omega^{2}E_{a}}{\omega_{b}^{2}%
}\frac{d}{d^{2}+g^{2}}\;.
\end{equation}
Using these results together with Eq. (\ref{1/tau GC}) one finds that%
\begin{align}
\frac{1}{\tau_{\varphi}} &  =2\gamma_{b}\left\vert \delta_{\alpha}\right\vert
^{2}\nonumber\\
&  \times\left[  \left(  1+\frac{4\Omega^{2}E_{a}}{\omega_{b}^{2}}\frac
{d}{d^{2}+g^{2}}\right)  \coth\frac{\beta\hbar\omega_{b}}{2}\right.
\nonumber\\
&  +\left.  \frac{\gamma_{a}}{\gamma_{b}}\frac{4\Omega^{2}E_{a}}{\omega
_{b}^{2}}\frac{1}{d^{2}+g^{2}}\coth\frac{\beta\hbar\omega_{a}}{2}\right]
\;.\nonumber\\
& \label{1/tau theta=0}%
\end{align}

The first term in Eq. (\ref{1/tau theta=0}) represents the contribution of the
thermal bath that is directly coupled to resonator $b$ to the dephasing rate.
This contribution can be either enhanced ($d>0$) or suppressed ($d<0$) due to
back-reaction effects. On the other hand, the last term in Eq.
(\ref{1/tau theta=0}) [compare with Eq. (71) of Ref. \cite{Buks_023815}]
represents the direct contribution of the driven resonator $a$. This
contribution can be understood in terms of the shift in the effective
resonance frequency of resonator $a$ between the two values corresponding to
the two coherent states $\left\vert \alpha_{1}\right\rangle $ and $\left\vert
\alpha_{2}\right\rangle $ (see Ref. \cite{Buks_023815}).

\section{Discussion}

We have considered above the case where $\Omega$ is small, $K_{a}=\gamma
_{a3}=0$ and $\gamma_{b}\ll\omega_{b}$. In addition, we have assumed that
$\gamma_{a}\ll\gamma_{b}$ in order to obtain the ISD of $X_{b}$, which is
given by Eq. (\ref{ISD gb ss ga}), and we have assumed the case $\theta=0$ to
obtain the dephasing rate, which is given by Eq. (\ref{1/tau theta=0}).
Furthermore, consider for simplicity the case of high temperature where
$\beta\hbar\omega_{b}\ll1$. For this case Eqs. (\ref{ISD gb ss ga}) and
(\ref{1/tau theta=0}) can be written in terms of the effective temperatures
$T_{\mathrm{ISD}}$ and $T_{\mathrm{D}}$%
\begin{equation}
\frac{1}{2\pi}\int\limits_{-\infty}^{\infty}\mathrm{d}\omega\;P_{X_{b}}\left(
\omega\right)  =\frac{2k_{\mathrm{B}}T_{\mathrm{ISD}}}{\hbar\omega_{b}}\ ,
\end{equation}%
\begin{equation}
\frac{1}{\tau_{\varphi}}=2\gamma_{b}\left\vert \delta_{\alpha}\right\vert
^{2}\frac{2k_{\mathrm{B}}T_{\mathrm{D}}}{\hbar\omega_{b}}\;,
\label{1/tau(T_D)}%
\end{equation}
where%
\begin{equation}
\frac{T_{\mathrm{ISD}}}{T}=1-\frac{\Omega^{2}E_{a}\Upsilon\left(  d,g\right)
}{\gamma_{a}\gamma_{b}}\left(  1-\frac{\omega_{b}\Theta_{a}}{\omega_{a}}%
\frac{\left(  1+d\right)  ^{2}+g^{2}}{3d}\right)  \;,
\end{equation}%
\begin{equation}
\frac{T_{\mathrm{D}}}{T}=1+\frac{4\Omega^{2}E_{a}}{\omega_{b}^{2}}\frac
{d}{d^{2}+g^{2}}\left(  1+\frac{\omega_{b}\gamma_{a}\Theta_{a}}{\omega
_{a}\gamma_{b}}\frac{1}{d}\right)  \;,
\end{equation}
and%
\begin{equation}
\Theta_{a}=\frac{\beta\hbar\omega_{a}}{2}\coth\frac{\beta\hbar\omega_{a}}%
{2}\;.
\end{equation}
In terms of $\left(  E_{a}\right)  _{c}$, which is given by Eq. (\ref{(E_a)_c}%
), one thus has%
\begin{equation}
\frac{T_{\mathrm{ISD}}}{T}=1-\frac{E_{a}}{\left(  E_{a}\right)  _{c}}%
\frac{\omega_{b}\Upsilon\left(  d,g\right)  }{\sqrt{3}\gamma_{b}}\left(
1-\frac{\omega_{b}\Theta_{a}}{\omega_{a}}\frac{\left(  1+d\right)  ^{2}+g^{2}%
}{3d}\right)  \;, \label{T_ISD}%
\end{equation}
and%
\begin{equation}
\frac{T_{\mathrm{D}}}{T}=1+\frac{4E_{a}}{\sqrt{3}\left(  E_{a}\right)  _{c}%
}\frac{dg}{d^{2}+g^{2}}\left(  1+\frac{\omega_{b}\gamma_{a}\Theta_{a}}%
{\omega_{a}\gamma_{b}}\frac{1}{d}\right)  \;. \label{T_D}%
\end{equation}
These results are valid only to lowest order in $\Omega$, however they may be
used in some cases to roughly estimate the lowest possible values of
$T_{\mathrm{ISD}}$ and $T_{\mathrm{D}}$. As can be seen from Eqs.
(\ref{T_ISD}) and (\ref{T_D}), the effective temperatures $T_{\mathrm{ISD}}$
and $T_{\mathrm{D}}$ may take considerably different values. This fact should
not be considered as surprising since the system is far from thermal
equilibrium and since the underlying mechanisms responsible for ISD reduction
and for suppression of decoherence are entirely different. In what follows, we
choose the parameters $d$ and $g$ such that the largest reduction in effective
temperature is achieved for a given $E_{a}$, and use these values to estimate
the lowest possible effective temperatures.

\subsection{Optimum ISD Reduction}

For the case of ISD reduction, we consider the case where the term that is
proportional to $\Theta_{a}$ in Eq. (\ref{T_ISD}), namely the term which
represents the contribution of the thermal baths that are directly coupled to
resonator $a$, is relatively small, namely the case where $\omega_{b}%
\Theta_{a}\ll\omega_{a}$. This condition is expected to be fulfilled for the
typical experimental situation. Most efficient ISD reduction is achieved by
choosing the parameters $g\ll1$ and $d=1$, for which the term $\Upsilon\left(
d,g\right)  $ obtains its maximum possible value $\Upsilon=1$ (see Fig.
\ref{Ups}). For this case Eq. (\ref{T_ISD}) becomes%
\begin{equation}
\frac{T_{\mathrm{ISD}}}{T}=1-\frac{\omega_{b}}{\sqrt{3}\gamma_{b}}\frac{E_{a}%
}{\left(  E_{a}\right)  _{c}}\left(  1-\frac{4\omega_{b}\Theta_{a}}%
{3\omega_{a}}\right)  \;. \label{T_ISD opt}%
\end{equation}
By taking%
\begin{equation}
E_{a}=\frac{\sqrt{3}\gamma_{b}}{\omega_{b}}\left(  E_{a}\right)  _{c}%
\equiv\left(  E_{a}\right)  _{\mathrm{ISD}}\;, \label{E_a _ISD}%
\end{equation}
Eq. (\ref{T_ISD opt}) yields the lowest possible value of $T_{\mathrm{ISD}}$,
which is denoted as $\left(  T_{\mathrm{ISD}}\right)  _{\min}$%
\begin{equation}
\frac{\left(  T_{\mathrm{ISD}}\right)  _{\min}}{T}=\frac{4\omega_{b}\Theta
_{a}}{3\omega_{a}}\;. \label{T_ISD min}%
\end{equation}

As was mentioned above, the above discussion is based on the approximated
result Eq. (\ref{T_ISD}), which expresses $T_{\mathrm{ISD}}$ to lowest
nonvanishing order in $\Omega$. Such an expansion apparently suggests that the
noise contribution due to the thermal bath that is directly coupled to
resonator $b$ can be altogether eliminated, leaving thus only the noise
contribution of the thermal baths that are directly coupled to resonator $a$
as a lower bound imposed upon $T_{\mathrm{ISD}}$ [see Eq. (\ref{T_ISD min})].
Obviously, however, higher orders in $\Omega$ have to be taken into account in
order estimate $\left(  T_{\mathrm{ISD}}\right)  _{\min}$ more accurately, as
was done in Ref. \cite{Blencowe_014511}, where $T_{\mathrm{ISD}}$ was expanded
up to forth order in $\Omega$.

\subsection{Decoherence Suppression}

For the case of decoherence suppression, on the other hand, the term that is
proportional to $\Theta_{a}$ in Eq. (\ref{T_D}) is not necessarily small for
the common experimental situation. We therefore chose the optimum values of
the parameters $d$ and $g$ for the more general case. Using the notation%
\begin{equation}
D=\frac{\omega_{b}\gamma_{a}\Theta_{a}}{\omega_{a}\gamma_{b}}\;,
\end{equation}
Eq. (\ref{T_D}) reads%
\begin{equation}
\frac{T_{\mathrm{D}}}{T}=1+\frac{4E_{a}}{\sqrt{3}\left(  E_{a}\right)  _{c}%
}f\left(  d,g,D\right)  \;.
\end{equation}
where%
\begin{equation}
f\left(  d,g,D\right)  =\frac{dg}{d^{2}+g^{2}}\left(  1+\frac{D}{d}\right)
\;.
\end{equation}
In general, the minimum value of the function $f\left(  d,g,D\right)  $ for a
given $g>0$ and a given $D>0$ is obtained at%
\begin{equation}
d_{\mathrm{m}}=-D-\sqrt{D^{2}+g^{2}}\;,
\end{equation}
and the minimum value is given by%
\begin{equation}
f\left(  d_{\mathrm{m}},g,D\right)  =-\frac{1}{2}\tan\left(  \frac{\tan
^{-1}\frac{g}{D}}{2}\right)  \;.
\end{equation}
The lowest value of $f\left(  d_{\mathrm{m}},g,D\right)  $ is thus obtained in
the limit $D\ll g$, for which one finds that $d_{\mathrm{m}}=-g$ and $f\left(
d_{\mathrm{m}},g,D\right)  =-1/2$. Therefore, one concludes that the largest
reduction in $T_{\mathrm{D}}$ for a given $E_{a}$ is obtained when%
\begin{equation}
\frac{\omega_{b}^{2}\Theta_{a}}{\omega_{a}\gamma_{b}}\ll1\;
\end{equation}
and when $d=-g$. For this case Eq. (\ref{T_D}) becomes%
\begin{equation}
\frac{T_{\mathrm{D}}}{T}=1-\frac{2}{\sqrt{3}}\frac{E_{a}}{\left(
E_{a}\right)  _{c}}\;. \label{T_D opt}%
\end{equation}
This results indicates that even when all parameters are optimally chosen such
that the largest reduction in $T_{\mathrm{D}}$ is obtained for a given $E_{a}%
$, no significant reduction in $T_{\mathrm{D}}$ is possible unless $E_{a}$
becomes comparable with $\left(  E_{a}\right)  _{c}$. Note, however, that in
our analysis of the present case the effect of nonlinear bistability has been
disregarded. This approximation can be justified for the case of ISD reduction
since, as can be seen from Eq. (\ref{E_a _ISD}), optimum reduction of the ISD
can be achieved well below the bistability threshold provided that $\gamma
_{b}\ll\omega_{b}$. On the other hand, Eq. (\ref{T_D opt}) indicates that
optimum suppression of decoherence can be achieved only very close to the
bistability threshold. In this region, however, our approximated treatment
breaks down and Eq. (\ref{T_D}) becomes inaccurate.

\begin{figure}
[ptb]
\begin{center}
\includegraphics[
height=3.6815in,
width=3.4411in
]%
{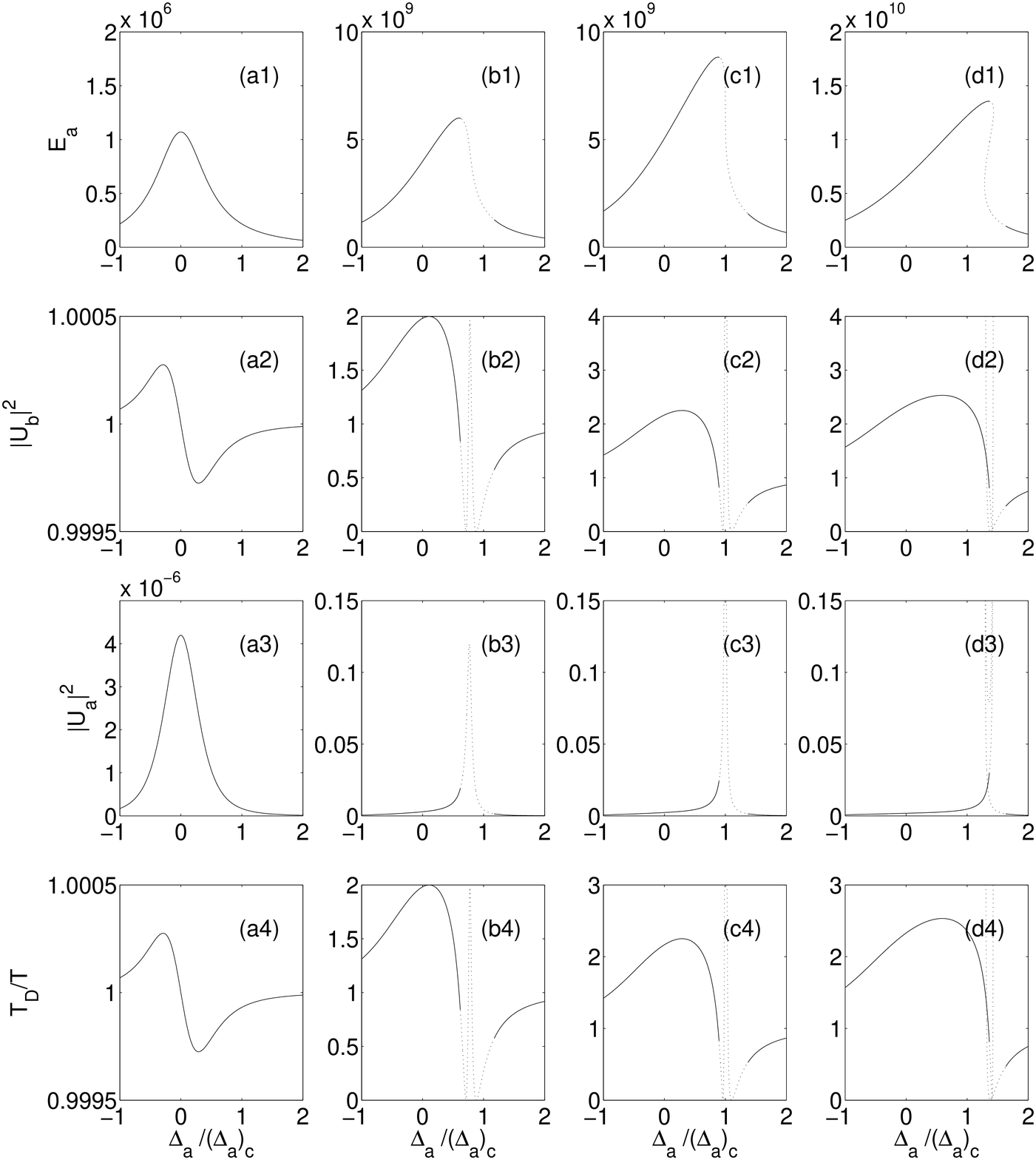}%
\caption{The factors $\left\vert U_{b}\right\vert ^{2}$ and $\left\vert
U_{a}\right\vert ^{2}$ and the ratio $T_{\mathrm{D}}/T$. The driving
amplitudes in columns \textrm{a}, \textrm{b}, \textrm{c} and \textrm{d} are
$b_{p}/\left(  b_{p}\right)  _{c}=0.01$, $0.8$, $1$ and $1.3$ respectively.
Other system parameters are $\Omega/\omega_{a}=10^{-10}$, $\omega_{b}%
/\omega_{a}=10^{-6}$, $\gamma_{b}/\omega_{b}=10^{-3}$, $K_{a}=2\times
2\Omega^{2}\omega_{b}/\left(  \omega_{b}^{2}+\gamma_{b}^{2}\right)  $,
$\theta=0$, $\gamma_{a1}/\omega_{b}=10^{2}$, $\gamma_{a2}/\gamma_{a1}=10^{-2}$
and $\gamma_{a3}=0.1\times\left\vert K_{a}^{\mathrm{eff}}\right\vert /\sqrt
{3}$. The ratio $T_{\mathrm{D}}/T$, which is plotted in the forth row, is
shown for the case $\beta\hbar\omega_{a}\ll1$. The dotted sections indicate
instability. }%
\label{FigEx1}%
\end{center}
\end{figure}

To calculate $T_{\mathrm{D}}$ near the bistability threshold we thus
numerically evaluate the dephasing rate given by Eq. (\ref{1/tau GC}) without
assuming that $\Omega$ is small or $K_{a}=\gamma_{a3}=0$. As before, we take
$\theta=0$ and consider for simplicity the case where $\beta\hbar\omega_{b}%
\ll1$, for which the effective temperature $T_{\mathrm{D}}$ is given by%
\begin{equation}
\frac{T_{\mathrm{D}}}{T}=\left\vert U_{b}\right\vert ^{2}+\frac{\Gamma
_{a}\omega_{b}\left\vert U_{a}\right\vert ^{2}}{\omega_{a}\gamma_{b}}%
\Theta_{a}\;.
\end{equation}
Figure \ref{FigEx1} shows an example calculation of the parameters $\left\vert
U_{b}\right\vert ^{2}$ and $\left\vert U_{a}\right\vert ^{2}$ and the ratio
$T_{\mathrm{D}}/T$ near bistability threshold of the system. The ratio
$T_{\mathrm{D}}/T$ is shown for the case where $\beta\hbar\omega_{a}\ll1$. The
set of system's parameters chosen for this example is listed in the caption of
Fig. \ref{FigEx1}. The stability of the mean filed solution is checked by
evaluating the eigenvalues of the matrix $W$. The dotted sections of the curve
$T_{\mathrm{D}}/T$ indicate the regions in which the solution is unstable
(where at least one of the eigenvalues of $W$ has a negative real part). Near
the onset of bistability point [see panel (c4) of Fig. \ref{FigEx1}] and near
jump points in the region of bistability [see panel (d4) of Fig. \ref{FigEx1}]
the ratio $T_{\mathrm{D}}/T$ may become relatively small. This behavior can be
attributed to critical slowing down, which occurs near these instability
points \cite{Buks_023815}. On the other hand, in the vicinity of these points
the solution becomes unstable [see the dotted sections of the curve
$T_{\mathrm{D}}/T$ in panels (c4) and (d4) of Fig. \ref{FigEx1}]. When the
unstable region is excluded one finds that no significant reduction in the
ratio $T_{\mathrm{D}}/T$ can be achieved for this particular example (the
lowest value is about $0.5$).%

In the previous example the mean-field solutions become unstable close to the
onset of bistability. This behavior prevents any significant suppression of
decoherence, namely, the ratio $T_{\mathrm{D}}/T$ could not be made much
smaller than unity. To overcome this limitation the parameter $\left(
E_{a}\right)  _{c}$, which is given by Eq. (\ref{|B|_a,c}), has to be
increased without, however, increasing the coupling parameter $\Omega$. We
point out below two possibilities to achieve this. In the first one, the
parameter $K_{a}$ is chosen such that $K_{a}\simeq2\Omega^{2}\omega
_{b}/\left(  \omega_{b}^{2}+\gamma_{b}^{2}\right)  ,$ and consequently
$K_{a}^{\mathrm{eff}}$ becomes very small [see Eq. (\ref{K_a^eff})]. In the
second one, which is demonstrated in Fig. \ref{FigEx2} below, the nonlinear
damping rate $\gamma_{a3}$ is chosen very close to the largest possible value
of $\left\vert K_{a}^{\mathrm{eff}}\right\vert /\sqrt{3}$ for which
bistability is accessible [see inequality (\ref{|K_a^eff|>})]. As can be see
from Eq. (\ref{|B|_a,c}), both possibilities allow significantly increasing
the parameter $\left(  E_{a}\right)  _{c}$. For the example shown in Fig.
\ref{FigEx2} below, the value $\gamma_{a3}=0.99\left\vert K_{a}^{\mathrm{eff}%
}\right\vert /\sqrt{3}$ is chosen and all other parameters are the same as in
the previous example (see caption of Fig. \ref{FigEx1}). As can be seen from
panels (c4) and (d4) of Fig. \ref{FigEx2}, much lower values of the ratio
$T_{\mathrm{D}}/T$ are achievable in the present example (a lowest value of
about $0.02$ is obtained at the edge of the region where the solution is
stable). This improvement can be attributed to the stabilization effect of the
nonlinear damping. It is important to point out, however, that implementation
of any of the two above mentioned possibilities require that the nonlinear
parameters of resonator $a$ ($K_{a}$ and/or $\gamma_{a3}$) can be accurately
tuned to the desired values. Such tuning of nonlinear parameters can possibly
becomes achievable by exploiting effects arising from thermo-optomechanical
coupling \cite{Zaitsev_1104_2237}.%

\begin{figure}
[ptb]
\begin{center}
\includegraphics[
height=3.6841in,
width=3.4411in
]%
{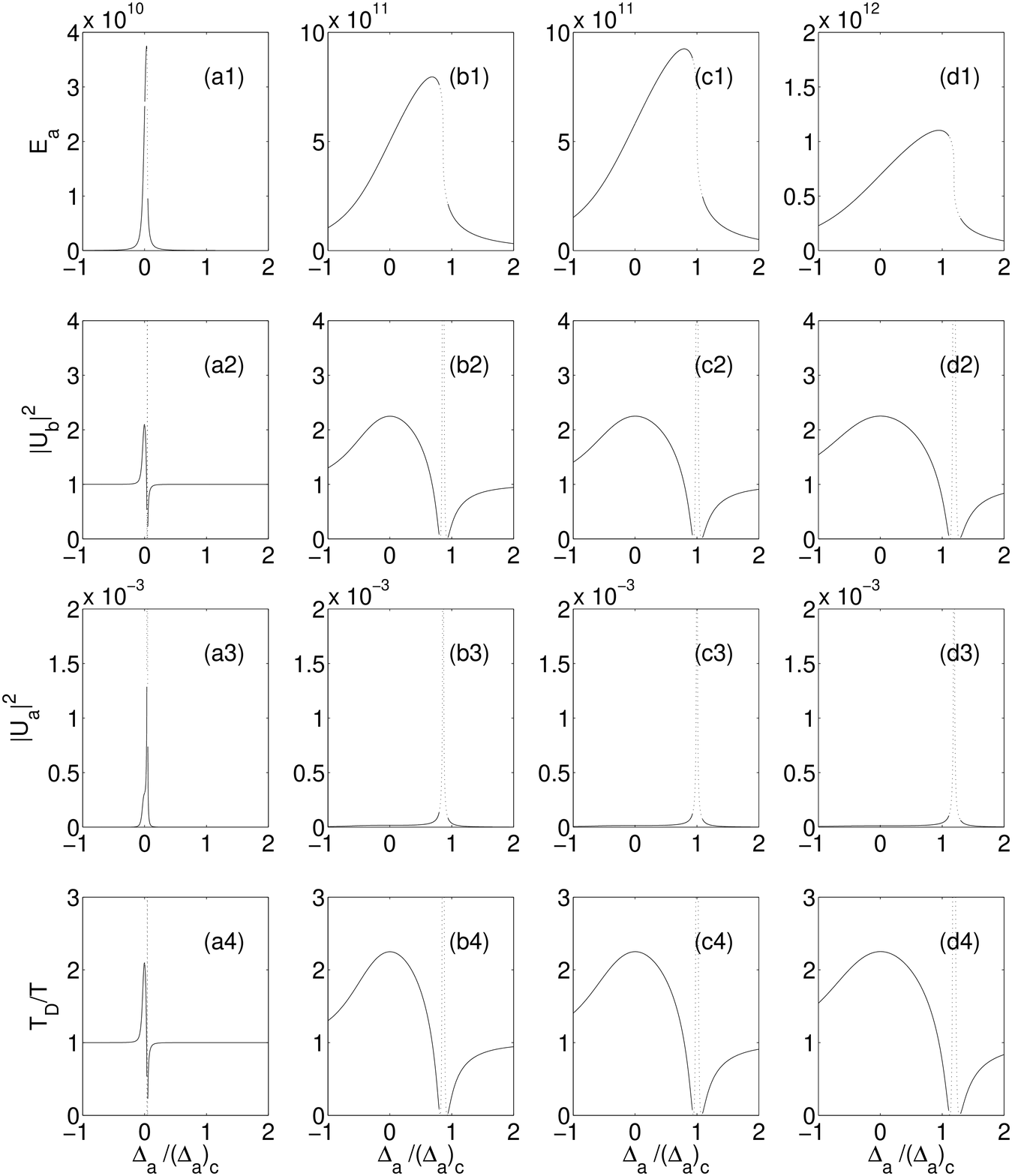}%
\caption{The factors $\left\vert U_{b}\right\vert ^{2}$ and $\left\vert
U_{a}\right\vert ^{2}$ and the ratio $T_{\mathrm{D}}/T$. In this example
$\gamma_{a3}=0.99\times\left\vert K_{a}^{\mathrm{eff}}\right\vert /\sqrt{3}$
whereas all other parameters are the same as in the previous example [see
caption of Fig. (\ref{FigEx1})].}%
\label{FigEx2}%
\end{center}
\end{figure}


\section*{Acknowledgment}

This work is supported by the German Israel Foundation under grant
1-2038.1114.07, the Israel Science Foundation under grant 1380021 and the
European STREP QNEMS Project.

\newpage
\bibliographystyle{ieee}
\bibliography{acompat,Eyal_Bib}

\newif\ifabfull\abfulltrue
\begin{thebibliography}{10}

\bibitem{Blencowe_159}
Miles Blencowe,
\newblock ``Quantum electromechanical systems,''
\newblock {\em Phys. Rep.}, vol. 395, pp. 159--222, 2004.

\bibitem{Schwab_36}
Keith~C. Schwab and Michael~L. Roukes,
\newblock ``Putting mechanics into quantum mechanics,''
\newblock {\em Phys. Today}, vol. July, pp. 36--42, 2005.

\bibitem{OConnell_697}
A.~D. O'Connell, M.~Hofheinz, M.~Ansmann, Radoslaw~C. Bialczak, M.~Lenander,
  Erik Luceroand~M. Neeley, D.~Sank, H.~Wang, M.~Weides, J.~Wenner, John~M.
  Martinis, and A.~N. Cleland,
\newblock ``Quantum ground state and single-phonon control of a mechanical
  resonator,''
\newblock {\em Nature}, vol. 464, pp. 697--703, 2010.

\bibitem{Legget_R415}
A.~J. Leggett,
\newblock ``Testing the limits of quantum mechanics: Motivation, state of play,
  prospects,''
\newblock {\em J. Phys. Condens. Matter}, vol. 14, pp. R415, 2002.

\bibitem{Leggett_857}
A.~J. Leggett and Anupam Garg,
\newblock ``Quantum mechanics versus macroscopic realism: Is the flux there
  when nobody looks?,''
\newblock {\em Phys. Rev. Lett.}, vol. 54, pp. 857--860, 1985.

\bibitem{Penrose_581}
Roger Penrose,
\newblock ``On gravity's role in quantum state reduction,''
\newblock {\em Gen. Relativ. Gravit.}, vol. 28, pp. 581--600, 1996.

\bibitem{Diosi_1165}
L.~Diosi,
\newblock ``Models for universal reduction of macroscopic quantum
  fluctuations,''
\newblock {\em Phys. Rev. A}, vol. 40, pp. 1165--1174, 1989.

\bibitem{Bose_4175}
S.~Bose, K.~Jacobs, and P.~L. Knight,
\newblock ``Preparation of nonclassical states in cavities with a moving
  mirror,''
\newblock {\em Phys. Rev. A}, vol. 56, pp. 4175, 1997.

\bibitem{Bose_3204}
S.~Bose, K.~Jacobs, and P.~L. Knight,
\newblock ``Scheme to probe the decoherence of a macroscopic object,''
\newblock {\em Phys. Rev. A}, vol. 59, pp. 3204--3210, 1999.

\bibitem{Kleckner_095020}
Dustin Kleckner, Igor Pikovski, Evan Jeffrey, Luuk Ament, Eric Eliel, Jeroen
  Van~Den Brink, and Dirk Bouwmeester,
\newblock ``Creating and verifying a quantum superposition in a
  micro-optomechanical system,''
\newblock {\em New J. Phys.}, vol. 10, pp. 095020, 2008.

\bibitem{Zurek_0306072}
Wojciech~H. Zurek,
\newblock ``Decoherence and the transition from quantum to classical --
  {REVISITED},''
\newblock {\em arXiv:quant-ph/0306072}, 2003.

\bibitem{Zurek_715}
Wojciech~Hubert Zurek,
\newblock ``Decoherence, einselection, and the quantum origins of the
  classical,''
\newblock {\em Rev. Mod. Phys.}, vol. 75, pp. 715--775, 2003.

\bibitem{Caldeira_587}
A.~O. Caldeira and A.~J. Leggett,
\newblock ``Path integral approach to quantum brownian motion,''
\newblock {\em Physica A}, vol. 121, pp. 587, 1983.

\bibitem{Joos_223}
E.~Joos and H.~D. Zeh,
\newblock ``The emergence of classical properties through interaction with the
  environment,''
\newblock {\em Physik B}, vol. 59, pp. 223, 1985.

\bibitem{Unruh_1071}
W.~G. Unruh and W.~H. Zurek,
\newblock ``Reduction of a wave packet in quantum brownian motion,''
\newblock {\em Phys. Rev. D}, vol. 40, pp. 1071, 1989.

\bibitem{Zurek_36}
W.~H. Zurek,
\newblock ``Decoherence and the transition from quantum to classical,''
\newblock {\em Physics Today}, vol. 44, pp. 36, 1991.

\bibitem{Rugar_699}
D.~Rugar and P.~Grutter,
\newblock ``Mechanical parametric amplification and thermomechanical noise
  squeezing,''
\newblock {\em Phys. Rev. Lett.}, vol. 67, pp. 699, 1991.

\bibitem{Almog_078103}
R.~Almog, S.~Zaitsev, O.~Shtempluck, and E.~Buks,
\newblock ``Noise squeezing in a nanomechanical duffing resonator,''
\newblock {\em Phys. Rev. Lett.}, vol. 98, pp. 78103, 2007.

\bibitem{Kimble_et_al_01}
H.~J. Kimble, Y.~Levin, A.~B. Matsko, K.~S. Thorne, and S.~P. Vyatchanin,
\newblock ``Conversion of conventional gravitational-wave interferometers into
  quantum nondemolition interferometers by modifying their input and/or output
  optics,''
\newblock {\em Phys. Rev. D}, vol. 65, pp. 022002, Dec 2001.

\bibitem{Braginsky_2002}
V.~B. Braginsky and S.~P. Vyatchanin,
\newblock ``Low quantum noise tranquilizer for {Fabry–Perot} interferometer,''
\newblock {\em Phys. Lett. A}, vol. 293, pp. 228--234, 2002.

\bibitem{Martin_125339}
Ivar Martin, Alexander Shnirman, Lin Tian, and Peter Zoller,
\newblock ``Ground-state cooling of mechanical resonators,''
\newblock {\em Phys. Rev. B}, vol. 69, pp. 125339, 2004.

\bibitem{Wilson-Rae_075507}
I.~Wilson-Rae, P.~Zoller, and A.~Imamolu,
\newblock ``Laser cooling of a nanomechanical resonator mode to its quantum
  ground state,''
\newblock {\em Phys. Rev. Lett.}, vol. 92, pp. 75507, 2004.

\bibitem{Clerk_238}
Aashish~A Clerk and Steven Bennett,
\newblock ``Quantum nanoelectromechanics with electrons, quasi-particles and
  cooper pairs: Effective bath descriptions and strong feedback effects,''
\newblock {\em New J. Phys.}, vol. 7, pp. 238, 2005.

\bibitem{Blencowe_236}
M.~P. Blencowe, J.~Imbers, and A.~D. Armour,
\newblock ``Dynamics of a nanomechanical resonator coupled to a superconducting
  single-electron transistor,''
\newblock {\em New J. Phys.}, vol. 7, pp. 236, 2005.

\bibitem{Wineland_0606180}
D.~J. Wineland, J.~Britton, R.~J. Epstein, D.~Leibfried, R.~B. Blakestad,
  K.~Brown, J.~D. Jost, C.~Langer, R.~Ozeri, S.~Seidelin, and J.~Wesenberg,
\newblock ``Cantilever cooling with radio frequency circuits,''
\newblock {\em arXiv: quant-ph/0606180}, 2006.

\bibitem{Marquardt_093902}
Florian Marquardt, Joe~P. Chen, A.~A. Clerk, and S.~M. Girvin,
\newblock ``Quantum theory of cavity-assisted sideband cooling of mechanical
  motion,''
\newblock {\em Phys. Rev. Lett.}, vol. 99, pp. 93902, 2007.

\bibitem{Braginsky&Manukin_67}
V.~B. Braginsky and A.~B. Manukin,
\newblock ``Ponderomotive effects of electromagnetic radiation (in
  {R}ussian),''
\newblock {\em ZhETF}, vol. 52, pp. 986--989, 1967.

\bibitem{Braginsky_et_al_70}
V.~B. Braginsky, A.~B. Manukin, and M.~Yu. Tikhonov,
\newblock ``Investigation of dissipative ponderomotive effects of
  electromagnetic radiation (in {R}ussian),''
\newblock {\em ZhETF}, vol. 58, pp. 1550--1555, 1970.

\bibitem{Corbitt_et_al_06}
T.~Corbitt, D.~Ottaway, E.~Innerhofer, J.~Pelc, and N.~Mavalvala,
\newblock ``Measurement of radiation-pressure-induced optomechanical dynamics
  in a suspended {Fabry-Perot} cavity,''
\newblock {\em Phys. Rev. A}, vol. 74, pp. 021802, Aug 2006.

\bibitem{Kippenberg&Vahala_08}
T.~J. Kippenberg and K.~J. Vahala,
\newblock ``Cavity optomechanics: Back-action at the mesoscale,''
\newblock {\em Science}, vol. 321, no. 5893, pp. 1172--1176, Aug 2008.

\bibitem{Schliesser_et_al_08}
A.~Schliesser, R.~Riviere, G.~Anetsberger, O.~Arcizet, and T.~J. Kippenberg,
\newblock ``Resolved-sideband cooling of a micromechanical oscillator,''
\newblock {\em Nat. Phys.}, vol. 4, pp. 415--419, 2008.

\bibitem{Genes_et_al_08}
C.~Genes, D.~Vitali, P.~Tombesi, S.~Gigan, and M.~Aspelmeyer,
\newblock ``Ground-state cooling of a micromechanical oscillator: {C}omparing
  cold damping and cavity-assisted cooling schemes,''
\newblock {\em Phys. Rev. A}, vol. 77, pp. 033804, Mar 2008.

\bibitem{Teufel_et_al_10}
J.~D. Teufel, D.~Li, M.~S. Allman, K.~Cicak, A.~J. Sirois, J.~D. Whittaker, and
  R.~W. Simmonds,
\newblock ``Circuit cavity electromechanics in the strong coupling regime,''
\newblock {\em arXiv}, Nov 2010.

\bibitem{Gigan_67}
S.~Gigan, H.~R. B{\"o}hm, M.~Paternostro, F.~Blaser, J.~B. Hertzberg, K.~C.
  Schwab, D.~Bauerle, M.~Aspelmeyer, and A.Zeilinger,
\newblock ``Self cooling of a micromirror by radiation pressure,''
\newblock {\em Nature}, vol. 444, pp. 67--70, 2006.

\bibitem{Arcizet_71}
O.~Arcizet, P.~F.Cohadon, T.~Briant, M.~Pinard, and A.~Heidmann,
\newblock ``Radiation-pressure cooling and optomechanical instability of a
  micromirror,''
\newblock {\em Nature}, vol. 444, pp. 71--74, 2006.

\bibitem{Kleckner_75}
D.~Kleckner and D.~Bouwmeester,
\newblock ``Sub-kelvin optical cooling of a micromechanical resonator,''
\newblock {\em Nature}, vol. 444, pp. 75--78, 2006.

\bibitem{Corbitt_150802}
T.~Corbitt, Y.~Chen, E.~Innerhofer, H.~M{\"u}ller-Ebhardt, D.~Ottaway,
  H.~Rehbein, D.~Sigg, S.~Whitcomb, C.~Wipf, and N.~Mavalvala,
\newblock ``An all-optical trap for a gram-scale mirror,''
\newblock {\em Phys. Rev. Lett.}, vol. 98, pp. 150802, 2007.

\bibitem{Schliesser_243905}
A.~Schliesser, P.~Del'Haye, N.~Nooshi, K.~J. Vahala, and T.~J. Kippenberg,
\newblock ``Radiation pressure cooling of a micromechanical oscillator using
  dynamical backaction,''
\newblock {\em Phys. Rev. Lett.}, vol. 97, pp. 243905, 2006.

\bibitem{Harris_013107}
J.~G.~E. Harris, B.~M. Zwickl, and A.~M. Jayich,
\newblock ``Stable, mode-matched, medium-finesse optical cavity incorporating a
  microcantilever mirror: Optical characterization and laser cooling,''
\newblock {\em Rev. Sci. Instrum.}, vol. 78, pp. 13107, 2007.

\bibitem{Naik_193}
A.~Naik, O.~Buu, M.~D. LaHaye, A.~D. Armour, A.~A. Clerk, M.~P. Blencowe, and
  K.~C. Schwab,
\newblock ``Cooling a nanomechanical resonator with quantum back-action,''
\newblock {\em Nature}, vol. 443, pp. 193--196, 2006.

\bibitem{Hohberger-Metzger_1002}
Constanze {H{\"o}hberger Metzger} and Khaled Karrai,
\newblock ``Cavity cooling of a microlever,''
\newblock {\em Nature}, vol. 432, pp. 1002--1005, 2004.

\bibitem{Teufel_1103_2144}
J.~D. Teufel, T.~Donner, Dale Li, J.~H. Harlow, M.~S. Allman, K.~Cicak, A.~J.
  Sirois, J.~D. Whittaker, K.~W. Lehnert, and R.~W. Simmonds,
\newblock ``Sideband cooling micromechanical motion to the quantum ground
  state,''
\newblock {\em arXiv:1103.2144}, 2011.

\bibitem{Yurke_5054}
Bernard Yurke and Eyal Buks,
\newblock ``Performance of cavity-parametric amplifiers, employing kerr
  nonlinearites, in the presence of two-photon loss,''
\newblock {\em J. Lightwave Tech.}, vol. 24, pp. 5054--5066, 2006.

\bibitem{Gardiner_3761}
C.~W. Gardiner and M.~J. Collett,
\newblock ``Input and output in damped quantum systems: Quantum stochastic
  differential equations and the master equation,''
\newblock {\em Phys. Rev. A}, vol. 31, pp. 3761, 1985.

\bibitem{Levinson_299}
Y.~Levinson,
\newblock ``Dephasing in a quantum dot due to coupling with a quantum point
  contact,''
\newblock {\em Europhys. Lett.}, vol. 39, pp. 299--304, 1997.

\bibitem{Buks_174504}
Eyal Buks and M.~P. Blencowe,
\newblock ``Decoherence and recoherence in a vibrating {RF} {SQUID},''
\newblock {\em Phys. Rev. B}, vol. 74, pp. 174504, 2006.

\bibitem{Buks_023815}
Eyal Buks and Bernard Yurke,
\newblock ``Dephasing due to intermode coupling in superconducting stripline
  resonators,''
\newblock {\em Phys. Rev. A}, vol. 73, pp. 23815, 2006.

\bibitem{Blencowe_014511}
M.~P. Blencowe and E.~Buks,
\newblock ``Quantum analysis of a linear {DC} {SQUID} mechanical displacement
  detector,''
\newblock {\em Phys. Rev. B}, vol. 76, pp. 14511, 2007.

\bibitem{Zaitsev_1104_2237}
Stav Zaitsev, Ashok~K. Pandey, Oleg Shtempluck, and Eyal Buks,
\newblock ``Forced and self oscillations of optomechanical cavity,''
\newblock {\em arXiv:1104.2237}, 2011.

\end{thebibliography}

\end{document}